# LIKELY VALUES OF THE COSMOLOGICAL CONSTANT

Hugo Martel[1,2], Paul R. Shapiro[1,3], and Steven Weinberg[4,5]

## ABSTRACT

In theories in which the cosmological constant takes a variety of values in different "subuniverses," the probability distribution of its observed values is conditioned by the requirement that there must be someone to measure it. This probability is proportional to the fraction of matter which is destined to condense out of the background into mass concentrations large enough to form observers. We calculate this "collapsed fraction" by a simple, pressure-free, spherically symmetric, nonlinear model for the growth of density fluctuations in a flat universe with arbitrary value of the cosmological constant, applied in a statistical way to the observed spectrum of density fluctuations at recombination. From this, the probability distribution for the vacuum energy density $\rho_V$ for Gaussian random density fluctuations is derived analytically. (The conventional quantity $\lambda_0$ is the vacuum energy density in units of the critical density at present, $\lambda_0 = \rho_V/\rho_{\mathrm{crit},0}$, where $\rho_{\mathrm{crit},0} = 3H_0^2/8\pi G$.) It is shown that the results depend on only one quantity, $\sigma^3\bar{\rho}$, where $\sigma^2$ and $\bar{\rho}$ are the variance and mean value of the fluctuating matter density field at recombination, respectively. To calculate $\sigma$, we adopt the flat CDM model with nonzero cosmological constant and fix the amplitude and shape of the primordial power spectrum in accordance with data on cosmic microwave background anisotropy from the COBE satellite DMR experiment. A comparison of the results of this calculation of the likely values of $\rho_V$ with present observational bounds on the cosmological constant indicates that the small, positive value of $\rho_V$ (up to 3 times greater than the present cosmic mass density) suggested recently by several lines of evidence is a reasonably likely value to observe, even if all values of $\rho_V$ are equally likely *a priori*.

[1] Department of Astronomy, The University of Texas, Austin, TX 78712

[2] hugo@sagredo.as.utexas.edu

[3] shapiro@astro.as.utexas.edu

[4] Theory Group, Department of Physics, The University of Texas, Austin, TX 78712

[5] weinberg@utaphy.ph.utexas.edu

*Subject headings:* cosmology: theory — galaxies: formation

## 1. INTRODUCTION

Though the evidence is still equivocal, there are persistent hints that the vacuum energy density[6] $\rho_V$ is positive, and up to 3 times greater than the present cosmic mass density $\rho_0$.[7] From the point of view of fundamental physics, such a value seems absurd. Crude estimates indicate a value of $\rho_V$ some 120 orders of magnitude greater than $\rho_0$, and while it is hard enough to imagine any sort of symmetry or adjustment mechanism that could make $\rho_V$ vanish (for a litany of failed attempts, see Weinberg [1989]), it would be even more peculiar for fundamental physical theory to dictate a non-zero value for $\rho_V$ that happens to be comparable to the cosmic mass density $\rho_0$ at this particular moment in the history of the universe.

As far as we know, the only way to understand a value of $\rho_V$ comparable to $\rho_0$ is based on a weak form of the anthropic principle. In several current theories the cosmological constant does not have a fixed value, but takes a variety of values with varying probabilities. For instance, Hawking (1983, 1984) showed that the introduction of a three-form gauge field $A_{\mu\nu\lambda}$ yields a state vector for the universe that is a superposition of terms with different values for the cosmological constant. Coleman (1988a) subsequently showed that the effect of wormholes in quantum gravity is to make the state vector a superposition of terms, in which *any* coupling coefficient in the Lagrangian that is not fixed by symmetries takes all possible values.[8] Also, in chaotic inflation (Linde 1986, 1987, 1988) the observed big bang is just one of an infinite number of expanding regions, in each of which the various fields that affect the vacuum energy can take different values. For brevity we will refer to parts of the "universe" in which the cosmological constant takes different values, such as terms in the state vector, local bangs, or whatever, as *subuniverses.*

---

[6]By $\rho_V$ is meant the sum of a term $\Lambda/8\pi G$, where $\Lambda$ is the cosmological constant appearing in the Einstein field equations, plus the contribution to the vacuum energy density from quantum fluctuations. The conventional quantity $\lambda_0$ is the vacuum energy density in units of the critical density at present, $\lambda_0 = \rho_V/\rho_{\mathrm{crit},0}$, where $\rho_{\mathrm{crit},0} = 3H_0^2/8\pi G$.

[7]For a review and earlier references, see Ostriker & Steinhardt (1995). This conclusion has been recently challenged by preliminary results of measurements by Perlmutter et al. (1996) of redshifts and distances for distant Type Ia supernovae.

[8]Coleman (1988b) subsequently concluded that the probability distribution of $\rho_V$ is sharply peaked at zero, as had previously been argued by Hawking (1983, 1984) and Baum (1984). This conclusion has been challenged by Fischler et al. (1989), and it will be assumed here that there is no sharp peak at $\rho_V = 0$.

In any theory of this general sort the measured value of the vacuum energy density $\rho_V$ would be much smaller than the value expected on dimensional grounds in elementary particle physics, not because there is any physical principle that makes it small in all subuniverses, but because it is only in the subuniverses where it is small that there would be anyone to measure it. This paper will show how to calculate the probability distribution of the values of $\rho_V$ that would be observed under these circumstances.

An earlier paper (Weinberg 1987) pointed out that the anthropic limit on the value of $\rho_V$ for $\rho_V > 0$ arises from the requirement that $\rho_V$ should not be so large as to prevent the formation of galaxies. This paper suggested that this requirement implies a value of $\rho_V$ roughly comparable to the cosmic density of nonrelativistic matter at the time that the earliest galaxies form, because, if $\rho_V$ were much larger than this, then galaxies could not form and there would be no observers, while there did not seem to be any reason for $\rho_V$ to be much smaller than this. Since then galaxies have continued to be found at higher and higher redshifts, and hence at higher and higher values of the cosmic mass density, and it is becoming clear that such values of $\rho_V$ are already ruled out. A galaxy with redshift $z \approx 4$ was formed when the cosmic mass density was more than (and perhaps considerably more than) $(1 + z)^3 \approx 125$ times the present mass density, which is much greater than the observational upper limit on $\rho_V$.

On the other hand, it is much more likely that the value of $\rho_V$ in our subuniverse is comparable to the average or median value measured by astronomers in all subuniverses, rather than the anthropic upper bound, so that its value should be compared with the cosmic mass density at the time of formation of typical galaxies, rather than of the earliest galaxies.[9] Here we will present a detailed Bayesian analysis, which allows a calculation of the probability distribution of $\rho_V$ from a knowledge of the spectrum of density fluctuations at recombination. The results suggest a much smaller likely value of $\rho_V$ than the anthropic

---

[9]See Weinberg (1996). This is essentially the same as what was called the "principle of mediocrity" by Vilenkin (1995a, 1995b, 1995c, 1995d). Vilenkin did not undertake a detailed calculation of the probability distribution of the cosmological constant. A calculation of this sort was done by Efstathiou (1995), but it contained some errors (Vilenkin 1995d; Weinberg 1996). Efstathiou's calculation was done numerically, using linear perturbation theory and what is believed to be a realistic model of initial perturbations, while the calculation presented here is thoroughly nonlinear, but concentrates on a single spherically symmetric density fluctuation, so that it is possible to understand the results analytically.

upper bound, a value that may not be in conflict with present observational bounds.

In §2 we describe how to calculate the probability distribution for the vacuum energy density $\rho_V$ that would be observed in various subuniverses in which there is someone to observe it. For $\rho_V > 0$, this is simply related to the fraction of matter that condenses into galaxies. We evaluate this fraction and the resulting probability distribution in §3 under the assumption of Gaussian random density fluctuations in the cosmic mass density at the time of recombination. The results depend only on the standard deviation $\sigma$ of these fluctuations at the time of recombination. In §4 we calculate $\sigma$, adopting the cold dark matter model for the power spectrum of these density fluctuations, and assuming a flat universe with nonzero cosmological constant (sometimes referred to as the flat "ΛCDM" model). The amplitude of the density fluctuations is fixed by the data on cosmic microwave background anisotropy from the COBE satellite DMR experiment. The results of this calculation of the likely values of $\rho_V$ are presented in §5, and compared with the range of values of the cosmological constant allowed by current observational and theoretical constraints. A summary and conclusions are presented in §6.

## 2. THE PROBABILITY DISTRIBUTION

We assume that the *a priori* probability of a net vacuum energy density between $\rho_V$ and $\rho_V + d\rho_V$ is $\mathcal{P}(\rho_V)\,d\rho_V$, where $\mathcal{P}(\rho_V)$ is some smooth function of $\rho_V$, with no special behavior near $\rho_V = 0$. What we want is the probability distribution $\mathcal{P}_{\mathrm{obs}}(\rho_V)$ that a random observer in any subuniverse will measure values of $\rho_V$ in a given range. According to the principles of Bayesian statistics, this is given by

$$\mathcal{P}_{\mathrm{obs}}(\rho_V) = \frac{\mathcal{A}(\rho_V)\mathcal{P}(\rho_V)}{\int_0^\infty \mathcal{A}(\rho_V)\mathcal{P}(\rho_V)\,d\rho_V} \ , \tag{1}$$

where $\mathcal{A}(\rho_V)$ is the mean number of astronomers making independent measurements of the vacuum energy density in subuniverses with vacuum energy density $\rho_V$.[10]

In calculating the quantity (1), we note that the range of anthropically allowed values of $\rho_V$ is so much smaller than the energy densities typical of elementary particle physics, that, within this narrow range, we can take the *a priori* probability distribution $\mathcal{P}(\rho_V)$ to

---

[10] We have not thought through the problems associated with infinite subuniverses, where $\mathcal{A}$ if nonzero is infinite. Presumably in this case we should take $\mathcal{A}$ to be the number of independent astronomers per same fixed number of baryons.

be constant. [11] The value of this constant then cancels in equation (1), which becomes

$$\mathcal{P}_{\text{obs}}(\rho_V) = \frac{\mathcal{A}(\rho_V)}{\int_0^\infty \mathcal{A}(\rho_V)\, d\rho_V} \ . \tag{2}$$

The evolution of galaxies and astronomers depends on a variety of constants of nature other than $\rho_V$, and the values of these other constants in the various subuniverses may be correlated with the values of $\rho_V$, but the range of values of $\rho_V$ that are anthropically allowed is so small compared with the energy densities typical of elementary particle physics that, within this range, we can take all other fundamental constants to have fixed values, the values we observe in our subuniverse.[12] Also, once a fluctuation in the cosmic mass distribution undergoes gravitational condensation, its subsequent evolution is essentially independent of $\rho_V$, so the ratio of astronomers to mass in galaxies may be taken as independent of $\rho_V$. The number of astronomers $\mathcal{A}(\rho_V)$ who can measure $\rho_V$ in any subuniverse should therefore be proportional to the fraction $\mathcal{F}(\rho_V)$ of matter incorporated in galaxies, so that equation (2) may be written

$$\mathcal{P}_{\text{obs}}(\rho_V) = \frac{\mathcal{F}(\rho_V)}{\int_0^\infty \mathcal{F}(\rho_V)\, d\rho_V} \ . \tag{3}$$

To calculate $\mathcal{F}(\rho_V)$, we note that the spectrum of initial density fluctuations at recombination can be regarded as independent of $\rho_V$, because values of $\rho_V$ for which galaxy formation is possible are much smaller than the cosmic mass density at or before the time of recombination. Similarly, it is reasonable to suppose that the total amount of matter in a subuniverse in theories of chaotic inflation is independent of $\rho_V$ within the narrow range of values of $\rho_V$ that are anthropically allowed. Our problem is then to calculate the fraction $\mathcal{F}$ of matter that undergoes gravitational condensation into galaxies as a function of $\rho_V$, for fixed initial conditions at recombination. Ironically, while the tininess of observationally allowed values of $\rho_V$ creates the cosmological constant problem in the first place, it is the tininess of the range of anthropically allowed values of $\rho_V$ that offers the possibility of a realistic calculation of $\mathcal{P}_{\rm obs}(\rho_V)$.

---

[11]It might be asked why within this range the *a priori* distribution $\mathcal{P}(\rho_V)$ is not a power of $\rho_V$ rather than a constant? A power law would mean that there is something special about the value $\rho_V = 0$, since a power law distribution function would have to vanish or blow up there. But the essence of the cosmological constant problem is that we do not know of anything in fundamental physics that gives a special significance to the value $\rho_V = 0$, which requires a precise cancellation of a coupling coefficient in the Lagrangian by radiative corrections. (Analogously, although the probability distribution of temperatures in the Antarctic ice must vanish at zero degrees Kelvin and zero degrees Centigrade, since these are the limits of the range of temperatures in which water freezes, we would hardly expect it to vanish or diverge at zero degrees Fahrenheit.) The same argument applies to a power-law dependence on $\log \rho_V$. Of course we are not assuming that $\mathcal{P}(\rho_V)$ does not contain terms that vary as positive powers of $\rho_V$, but only that there is *also* a constant term, which then naturally dominates for the very small values of $\rho_V$ that are consistent with the appearance of astronomers who can measure $\rho_V$.

[12]If various constants of nature and initial conditions vary from one subuniverse to another independently of the values of $\rho_V$, then $\mathcal{P}_{\text{obs}}(\rho_V) d\rho_V$ is the probability that, *if* the other constants and initial conditions take the values we observe, then the vacuum energy density will be observed to be between $\rho_V$ and $\rho_V + d\rho_V$.

To see how this can work in practice, we will carry out an illustrative calculation of $\mathcal{F}(\rho_V)$ and use it to calculate both the integrated probability distribution and the mean and median values of $\rho_V$ observed by all astronomers in the subuniverses that contain astronomers. Earlier work (Weinberg 1987) used a very simple model (Peebles 1967) of galaxy formation from isolated spherically symmetric pressureless fluctuations. This calculation was improved in the report of a recent conference talk (Weinberg 1996), by using the well-known model of Gunn & Gott (1972), which also assumes isolated fluctuations with spherical symmetry and zero pressure, but includes the infall of matter from outside the initially overdense ball. Here we will also take into account the facts that, with space filled with fluctuations, there is a limit to the mass that can accrete onto any one fluctuation, and that there must be regions of negative as well as positive overdensity.

Consider a spherically symmetric pressureless fluctuation, consisting at recombination of a spherical core of volume $V$ and positive average fractional overdensity $\delta$ [i.e. $\delta \equiv (\rho - \bar{\rho})/\bar{\rho}$, where $\rho$ is the average density inside $V$ and $\bar{\rho}$ is the cosmic mean density at recombination], surrounded by a spherical shell of volume $U$ of constant fractional underdensity, taken to have the value $(V/U)\delta$, so that the average overdensity within the whole fluctuation is zero. Outside this shell are other fluctuations, about which we do not need to say anything, except to assume that they do not seriously interfere with the spherical symmetry of the fluctuation in question. For simplicity, we will take $V/U$ to be the same for all fluctuations.

As shown in earlier work (Weinberg 1987) the core will undergo gravitational collapse if[13]

$$\delta \geq \left(\frac{729\rho_V}{500\bar{\rho}}\right)^{1/3} . \tag{4}$$

In addition, that portion $U'$ of the outer shell will fall into the core, for which the average fractional overdensity within the volume $V + U'$ saturates this inequality:

$$\frac{V\delta - U'(V/U)\delta}{V + U'} = \left(\frac{729\rho_V}{500\bar{\rho}}\right)^{1/3} . \tag{5}$$

---

[13]It was originally assumed (Weinberg 1987) that the overdensity $\delta\rho$ was uniform, but these results actually hold for arbitrary spherically symmetric fluctuations, with $\delta$ interpreted as the *average* fractional overdensity.

The fraction of the total mass $\bar{\rho}(U + V)$ that suffers gravitational contraction will be

$$\mathcal{F}(\delta, \rho_V) = \frac{(1 + \delta) V + [1 - (V/U)\delta] U'}{(U + V)} \tag{6}$$

Solving equation (5) for $U'$, we find for the fraction of mass that undergoes gravitational contraction

$$\mathcal{F}(\delta, \rho_V) = (V/U)\delta \left[ \frac{1 + (729\rho_V/500\bar{\rho})^{1/3}}{(729\rho_V/500\bar{\rho})^{1/3} + (V/U)\delta} \right] \; . \tag{7}$$

We require that the total density be everywhere nonnegative, so that

$$\delta \leq (U/V) \; . \tag{8}$$

For $\delta$ satisfying this inequality, equation (7) gives $\mathcal{F} \leq 1$, so that no fluctuation can get more than its fair share of mass.

In what follows we will assume that the fluctuation number density $\mathcal{N}(\delta)$ is negligible for initial fluctuations that are not everywhere weak, so that we will be integrating only over fluctuations for which $\delta \ll 1$ *and* $\delta \ll U/V$. Also, for any anthropically allowed cosmological constant, $\rho_V$ is much less than the mass density $\bar{\rho}$ at recombination, so we will drop the term $(729\rho_V/500\bar{\rho})^{1/3}$ in the numerator (but not the denominator) of the fraction in square brackets in equation (7). The fraction of mass winding up in galaxies is then

$$
\begin{aligned}
\mathcal{F}(\rho_V) &= \int_{(729\rho_V/500\bar{\rho})^{1/3}}^{\infty} d\delta \; \mathcal{N}(\delta) \, \mathcal{F}(\delta, \rho_V) \\
&= \int_{(729\rho_V/500\bar{\rho})^{1/3}}^{\infty} d\delta \; \frac{(V/U)\delta \, \mathcal{N}(\delta)}{(729\rho_V/500\bar{\rho})^{1/3} + (V/U)\delta} \; ,
\end{aligned} \tag{9}
$$

where $\mathcal{N}(\delta)\, d\delta$ is the fraction of all positive fluctuations that have average core fractional overdensity between $\delta$ and $\delta + d\delta$, normalized so that

$$\int_0^{\infty} \mathcal{N}(\delta) \, d\delta = 1 \; .$$

The normalization integral in equation (3) can be calculated by interchanging the order of integration over $\delta$ and $\rho_V$, and expressing $\rho_V$ in terms of a dimensionless variable $x$ defined by

$$\rho_V = \frac{500 x^3 \bar{\rho}\, \delta^3}{729} \; .$$

Equation (9) then gives

$$\int_0^{\infty} \mathcal{F}(\rho_V) \, d\rho_V = \frac{500}{243}\bar{\rho} \, \langle \delta^3 \rangle \, (V/U) \, I_0(V/U) \; , \tag{10}$$

where $I_0(s)$ is the function

$$I_0(s) \equiv \int_0^1 \frac{x^2 \, dx}{x + s} = \frac{1}{2} - s - s^2 \ln\left(\frac{s}{1+s}\right) \;, \qquad (11)$$

and the brackets denote an average over all positive fluctuations

$$\langle f(\delta) \rangle \equiv \int_0^\infty \mathcal{N}(\delta) \, f(\delta) \, d\delta \;. \qquad (12)$$

The normalized probability distribution (3) for the observed vacuum energy density is then

$$\begin{aligned}
\mathcal{P}_{\text{obs}}(\rho_V) &= \frac{243}{500} \frac{1}{\langle \delta^3 \rangle I_0(V/U)} \\
&\times \int_{(729\rho_V/500\bar{\rho})^{1/3}}^\infty d\delta \; \frac{\delta \, \mathcal{N}(\delta)}{(729\rho_V/500\bar{\rho})^{1/3} + (V/U)\delta} \;,
\end{aligned} \qquad (13)$$

with all quantities on the right-hand side referring to the time of recombination.

In using equation (13) we will need to make some assumption about the shape parameter $s \equiv V/U$. The value $s = 0$ corresponds to the limit in which each positive fluctuation is isolated, surrounded by an infinite volume of compensating underdensity (at a total density arbitrarily close to its mean value $\bar{\rho}$), the case considered by Weinberg (1996). Values of $s$ much greater than unity correspond to the limit in which the additional mass associated with the compensating underdense volume $U$ is insignificant compared with that contained within the positive fluctuation in volume $V$, the case considered in Weinberg (1987). The value $s = 1$ corresponds to the case in which every positive fluctuation is surrounded by an *equal* volume of compensating negative fluctuation. This latter value is the one most relevant to a Gaussian-random distribution of linear density fluctuations, since the volumes occupied by positive and negative density fluctuations of equal amplitude are exactly equal in that case. Thus we will concentrate on the value $s = 1$ when we apply our analysis to the observed universe in what follows. Fortunately, as we shall see, most of our results will turn out to be almost independent of the value chosen for $s$.

Strictly speaking, equation (13) times $d\rho_V$ gives the probability that, *if the vacuum energy density is positive*, then it will be observed to be between $\rho_V$ and $\rho_V + d\rho_V$. For $\rho_V < 0$, the anthropic bound on $\rho_V$ is set by the condition (Barrow & Tipler 1986) that the subuniverse should survive long enough for intelligent life to arise. It is plausible that $\rho_V > 0$ is strongly favored anthropically (Weinberg 1996) as well as observationally, but we will make no attempt here to calculate the probability distribution for negative values of $\rho_V$.

Equation (13) can be used in various ways. One is to calculate the mean of various powers of $\rho_V$. Carrying out the same exchange of integrations and change of variables as in our calculation of equation (10), we find

$$< \rho_V^n > = \left( \frac{500}{729} \right)^n \frac{\bar{\rho}^n \langle \delta^{3n+3} \rangle}{\langle \delta^3 \rangle} \frac{I_n(V/U)}{I_0(V/U)} \ , \tag{14}$$

where $I_n$ is the function

$$I_n(s) \equiv \int_0^1 \frac{x^{3n+2} \ dx}{x + s} \ . \tag{15}$$

The average $< \rho_V^n >$ in equation (14) is taken over all subuniverses, not over fluctuations as in $\langle \delta \rangle$, and again all quantities on the right-hand side are to be evaluated at recombination. In particular, the mean value observed for $\rho_V$ (if $\rho_V$ is positive) is

$$< \rho_V > = \frac{500}{729} \frac{\langle \delta^6 \rangle \bar{\rho}}{\langle \delta^3 \rangle} \frac{I_1(V/U)}{I_0(V/U)} \ , \tag{16}$$

with

$$I_1(s) = \frac{1}{5} - \frac{s}{4} + \frac{s^2}{3} - \frac{s^3}{2} + s^4 + s^5 \ln \left( \frac{s}{1+s} \right) \ . \tag{17}$$

Fortunately the ratio $I_1(V/U)/I_0(V/U)$ in equation (16) turns out to be nearly constant; it drops from a value 0.5 when $s \equiv V/U \gg 1$, corresponding to no infall, to a value 0.4 when $s \ll 1$, corresponding to well separated fluctuations. It is therefore not very important what value we choose for $s$. As mentioned earlier, it seems reasonable to assume the intermediate case $s = 1$, where overdense and underdense regions have equal volume at recombination; in this case, the ratio $I_1/I_0$ takes the value

$$I_1(1)/I_0(1) = \frac{\frac{47}{60} - \ln 2}{-\frac{1}{2} + \ln 2} = 0.46693 \ .$$

The insensitivity of our results to the value of $s$ suggests that they also may not be much affected by the crudeness of our treatment of the effect of one fluctuation on another.

We could also use equation (13) to calculate the integrated probability $\mathcal{P}(> \rho_V)$ that the vacuum energy density is greater than $\rho_V$:

$$\mathcal{P}(> \rho_V) \equiv \int_{\rho_V}^{\infty} \mathcal{P}_{\mathrm{obs}}(\rho_V') \ d\rho_V' \ . \tag{18}$$

With the same reversal of integrations over $\rho_V'$ and $\delta$ and the same change of variables as before, we find

$$\mathcal{P}(> \rho_V) = \frac{\left\langle \delta^3 \ I\left( (729\rho_V/500\bar{\rho})^{1/3}/\delta \ , V/U \right) \right\rangle}{\langle \delta^3 \rangle \ I_0(V/U)} \ , \tag{19}$$

where for $t < 1$,

$$I(t,s) \equiv \int_t^1 \frac{x^2}{x+s} dx = \frac{1}{2}(1-t^2) - s(1-t) + s^2 \ln\left(\frac{1+s}{t+s}\right) \ , \tag{20}$$

and $I(t,s) \equiv 0$ for $t > 1$. As in our calculation of $<\rho_V>$, this is insensitive to the precise value of $s \equiv V/U$. For very small $\rho_V$, equation (19) of course approaches unity, whatever the value of $s$. For very large $\rho_V$, the only fluctuations that contribute within the integral over $\delta$ in equation (19) are those with $\delta$ near the lower limit of the integral, for which the lower limit of the integral (20) is near unity, where this integral behaves as $1/(1+s)$. But then $\mathcal{P}(> \rho_V)$ has an $s$-dependence proportional to $[I_0(s)(1+s)]^{-1}$, which only rises from 2 to 3 as $s$ rises from zero to infinity.

## 3.  Gaussian Density Fluctuations

To go further, we must make some assumption about the form of the fluctuation probability distribution $\mathcal{N}(\delta)$ at an early epoch, such as that of recombination. Current data on the anisotropy of the cosmic microwave background on large angular scales and on the large-scale clustering properties of galaxies, as well as theoretical predictions of the origin of density fluctuations by quantum processes in the early universe in inflationary cosmology models, are consistent with the assumption that the primordial fluctuations were isotropic, Gaussian random noise of very small amplitude. In this case the fluctuation distribution has the form

$$\mathcal{N}(\delta) = \frac{1}{\sigma}\left(\frac{2}{\pi}\right)^{1/2} \exp\left(-\frac{\delta^2}{2\sigma^2}\right) \ . \tag{21}$$

The mean values of powers of $\delta$ are given in terms of the variance $\sigma^2$ by

$$\langle \delta^N \rangle = \frac{2^{N/2}\sigma^N}{\pi^{1/2}} \, \Gamma\left(\frac{N+1}{2}\right) \ . \tag{22}$$

Equation (14) then gives

$$<\rho_V^n> = \Gamma\left(\frac{3n+4}{2}\right) \left(\frac{1000 \cdot 2^{1/2}\sigma^3\bar\rho}{729}\right)^n \frac{I_n(s)}{I_0(s)} \ , \tag{23}$$

and in particular, the mean vacuum energy density is

$$<\rho_V> = \left[\frac{625(2\pi)^{1/2}\sigma^3\bar\rho}{243}\right] \frac{I_1(s)}{I_0(s)} \ , \tag{24}$$

where, as before, $s \equiv V/U$. This gives the numerical values

$$< \rho_V > = \sigma^3 \bar{\rho} \times \left\{ \begin{array}{ll} 2.5788 & s = 0 \,, \\ 3.0103 & s = 1 \,, \\ 3.2235 & s = \infty \,. \end{array} \right. \tag{25}$$

Also, using the Gaussian distribution (21) in equation (19) and writing

$$\delta = \left( \frac{729 \rho_V}{500 \bar{\rho}} \right)^{1/3} \left( \frac{x}{\beta} \right)^{1/2} \,,$$

we find the differential probability distribution

$$\mathcal{P}_{\text{obs}}(\rho_V) \, d\rho_V = \frac{\beta^{1/2} d\beta}{2 I_0(s)} \int_\beta^\infty \frac{e^{-x} \, dx}{s x^{1/2} + \beta^{1/2}} \,, \tag{26}$$

where

$$\beta \equiv \frac{1}{2\sigma^2} \left( \frac{729 \rho_V}{500 \bar{\rho}} \right)^{2/3} \,. \tag{27}$$

The probability of a vacuum energy density greater than $\rho_V$ is then:

$$\mathcal{P}(>\rho_V) = (1 + \beta)e^{-\beta} - \frac{1}{2 I_0(s)} \int_\beta^\infty e^{-x} \left\{ -2s(\beta x)^{1/2} + \beta + 2s^2 x \ln \left[ \frac{1}{s} \left( \frac{\beta}{x} \right)^{1/2} + 1 \right] \right\} dx \,. \tag{28}$$

By combining equations (24) and (27), we see that the parameter $\beta$ in equations (26) and (28) may be expressed in terms of $s \equiv V/U$ and the ratio $\rho_V / <\rho_V>$ of the vacuum energy density to its mean value

$$\beta = \frac{\pi^{1/3}}{4} \left[ 15 \frac{\rho_V}{<\rho_V>} \frac{I_1(s)}{I_0(s)} \right]^{2/3} \,. \tag{29}$$

Thus the probability of observing a vacuum energy density in a certain range depends only on the values of $\rho_V / <\rho_V>$ and $s$. The analysis of initial fluctuations will enter here only as a means of calculating the parameter $\bar{\rho}\sigma^3$, and hence $<\rho_V>$.

In using these formulas, it is important to note that during the era of recombination, when $\rho_V$ is negligible and fluctuations are small, $\sigma$ grows as $t^{2/3}$ and $\bar{\rho}$ falls as $t^{-2}$, so the combination $\sigma^3 \bar{\rho}$ is time independent. Therefore equations (24)–(28) show that $<\rho_V>$, $\mathcal{P}_{\text{obs}}(\rho_V)$, and $P(>\rho_V)$ do not depend on what we take as the precise moment of recombination at which $\sigma$ and $\bar{\rho}$ are evaluated.

We have plotted the differential probability $d\mathcal{P}/d\alpha \equiv\ <\rho_V>\ \mathcal{P}_{\mathrm{obs}}(\rho_V)$ versus $\alpha \equiv \rho_V/<\rho_V>$ in Figure 1, for a range of values of $s$. This figure shows that the probability distribution drops exponentially for large $\rho_V$, and is remarkably insensitive to the value of $s$ for all $\rho_V$, except for $\rho_V \ll 1$. We have also plotted the differential probability per logarithmic interval of $\alpha$, $\alpha \mathcal{P}_{\mathrm{obs}}(\alpha)$, versus $\alpha$ [i.e. $d\mathcal{P}/d\log\alpha = \rho_V \mathcal{P}_{\mathrm{obs}}(\rho_V)$] in the bottom panel of Figure 1. This quantity is peaked at values of $\alpha \cong 0.7 - 0.8$, almost independent of $s$. Figure 2 shows the integrated probability $\mathcal{P}(> \rho_V)$ versus the dimensionless quantity $\beta$ defined by equation (27), for various values of $s$. Again, we see that this probability is only weakly dependent on $s$. We can evaluate $\mathcal{P}(> \rho_V)$ exactly in a few special cases

$$\mathcal{P}(> \rho_V) = \begin{cases} e^{-\beta} & s = 0, \\ \frac{3}{2}e^{-\beta} & s = \infty \ \ \& \ \ \beta \gg 1 \ . \end{cases} \tag{30}$$

Equation (28) may be used to calculate the median value $(\rho_V)_{1/2}$, for which $\mathcal{P}[> (\rho_V)_{1/2}] \equiv 1/2$. For isolated fluctuations $s = 0$, and the result is

$$(\rho_V)_{1/2} \ = \ \frac{4(\ln 2)^{3/2}}{3\pi^{1/2}} <\rho_V> = \ 0.43411 \ <\rho_V> = \ 1.1195 \ \sigma^3 \bar{\rho} \ . \tag{31}$$

For other values of $s$ the median must be calculated numerically, with the result that

$$(\rho_V)_{1/2} \ = <\rho_V> \times \begin{cases} 0.43411 \\ 0.49335 \\ 0.51480 \end{cases} = \sigma^3 \bar{\rho} \times \begin{cases} 1.1195 & s = 0, \\ 1.4851 & s = 1, \\ 1.6595 & s = \infty. \end{cases} \tag{32}$$

Again we see the insensitivity of our results to the shape parameter $s \equiv V/U$.

It is also interesting to ask what is the *range* of reasonably likely values of $\rho_V$? For instance, what is the range of values of $\rho_V$ for which only 5% of astronomers in all subuniverses would observe smaller values, and only 5% would observe larger values? By setting $\mathcal{P}(> \rho_V)$ in equation (28) equal to 0.95 and 0.05 and solving for $\rho_V$, we find that this range has the lower bound

$$(\rho_V)_{0.95} \ = <\rho_V> \times \begin{cases} 0.00874 \\ 0.01959 \\ 0.02359 \end{cases} = \sigma^3 \bar{\rho} \times \begin{cases} 0.02254 & s = 0, \\ 0.05897 & s = 1, \\ 0.07604 & s = \infty, \end{cases} \tag{33}$$

and the upper bound

$$(\rho_V)_{0.05} \ = <\rho_V> \times \begin{cases} 3.9005 \\ 3.6914 \\ 3.6157 \end{cases} = \sigma^3 \bar{\rho} \times \begin{cases} 10.0586 & s = 0, \\ 11.1122 & s = 1, \\ 11.6552 & s = \infty. \end{cases} \tag{34}$$

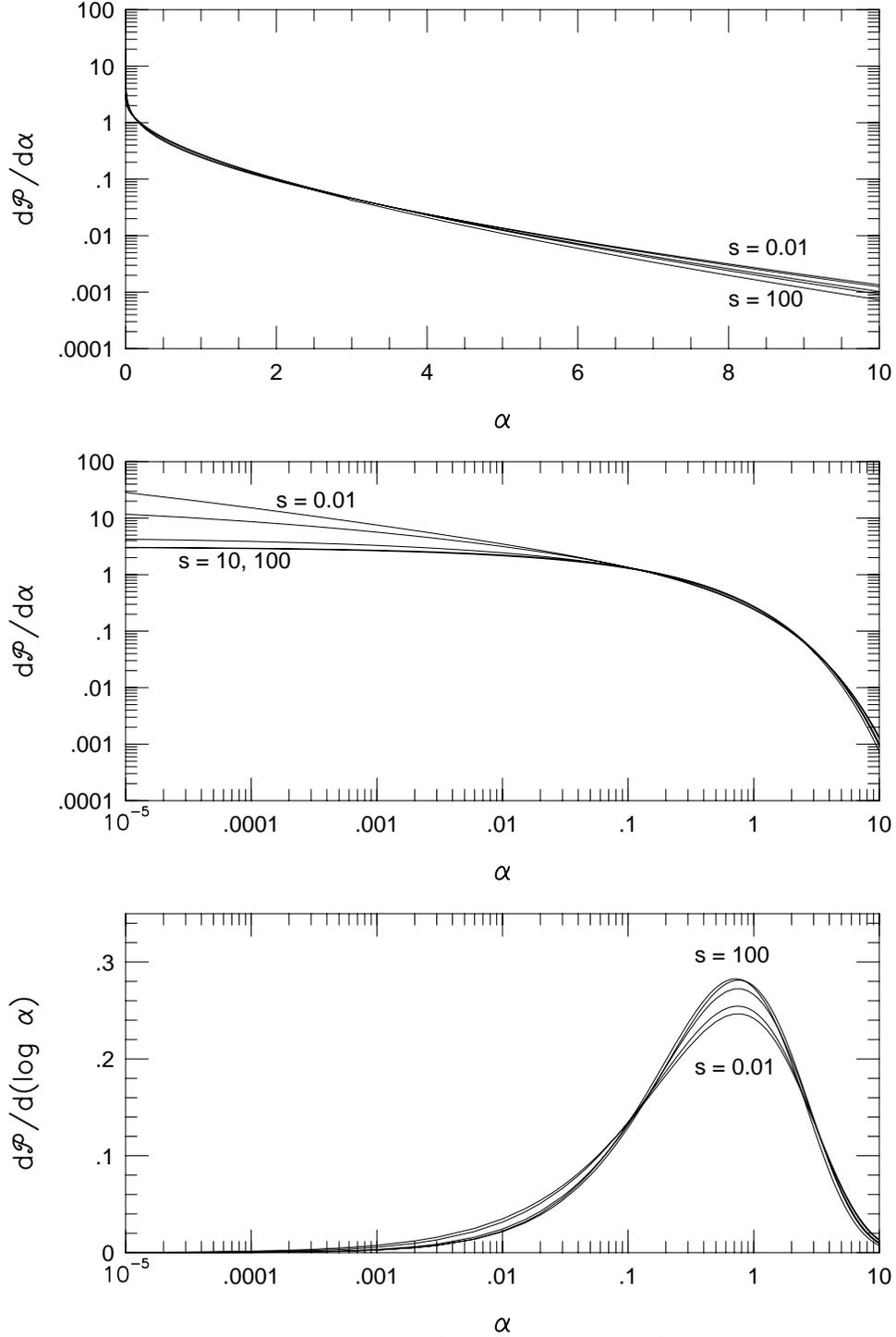

Fig. 1.— Differential probability $d\mathcal{P}/d\alpha$ $[=<\rho_V>\mathcal{P}_{\mathrm{obs}}(\rho_V)]$ versus $\alpha$, where $\alpha \equiv \rho_V/<\rho_V>$, for $s = 0.01$, $0.1$, $1$, $10$, and $100$. The bottom panel shows $d\mathcal{P}/d\log\alpha$ $[=\rho_V\mathcal{P}_{\mathrm{obs}}(\rho_V)]$ versus $\alpha$, instead.

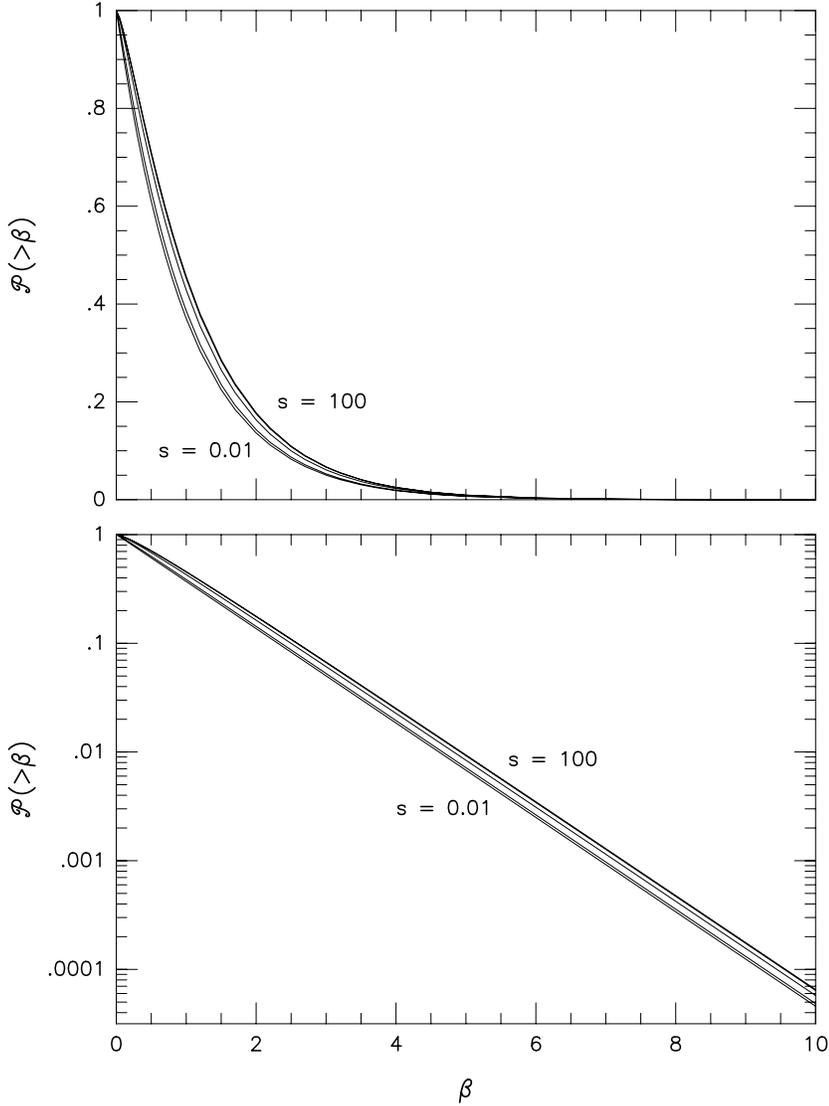

Fig. 2.— Integrated probability $\mathcal{P}(>\beta)$ versus $\beta$, for $s = 0.01$, $0.1$, $1$, $10$, and $100$.

Although the lower bound (33) is evidently somewhat sensitive to the shape parameter $s$, we see that for all values of $s$ the distribution of $\rho_V$ values is quite broad; it would not be very unlikely for a subuniverse to have a value of $\rho_V$ that is 50 times smaller or 3.7 times larger than the average. On the other hand, it would be extremely unlikely to observe a value of $\rho_V$ which differs from the mean by more than a few orders of magnitude. Not only are *large* values of $\rho_V$ unlikely, therefore, as we might previously have guessed based upon the fact that large $\rho_V$ suppresses galaxy formation, but values of $\rho_V$ extremely close to zero are also unlikely; there are simply too many other subuniverses to observe which have larger values of $\rho_V$ but not large enough to prevent galaxy formation.

For some purposes it is useful to have an analytic fit to the integrated probability. The following generally works quite well:

$$\mathcal{P}(> \rho_V) \approx \frac{(3s+2)e^{-\beta} - se^{-2\beta}}{2s+2} \ . \tag{35}$$

Notice that this fit is consistent with the special cases mentioned above. The absolute difference $|\mathcal{P} - f|$ and the relative difference $|(\mathcal{P} - f)/\mathcal{P}|$ are always less than 0.025 and 0.028, respectively, for $0 \leq s \leq 1$, while these errors increase to 0.027 and 0.092, respectively, for $1 \leq s \leq \infty$. Because the approximation formula (35) gives $\mathcal{P}(> \rho_V)$ as a quadratic function of $e^{-\beta}$, it is easy to solve equation (35) for the value of $\beta$ and hence of $\rho_V / < \rho_V >$ that give any specific value for $\mathcal{P}(> \rho_V)$. For instance, for $s = 1$ equation (35) yields a median vacuum density $(\rho_V)_{1/2} = 0.482 < \rho_V >$, as compared with an exact value $(\rho_V)_{1/2} = 0.493 < \rho_V >$, while for $s = \infty$ equation (35) yields $(\rho_V)_{1/2} = 0.568 < \rho_V >$, as compared with an exact value $(\rho_V)_{1/2} = 0.515 < \rho_V >$. For $s = 0$, equation (35) is exact.

Finally, it is also useful to compute the probability that $\alpha$ is observed to be within the interval $[\alpha_1, \alpha_2]$, according to $\mathcal{P}(\alpha_1 \leq \alpha \leq \alpha_2) \equiv \mathcal{P}(> \alpha_1) - \mathcal{P}(> \alpha_2)$. We present these interval probability results in Table 1, for $s = 1$. We also plot in Figure 3 the interval probability isocontours such that, for any given value of $\alpha$, call it $\alpha^*$, the curves show the values of $\alpha$ above and below this $\alpha^*$ at which the interval probability has some particular value, as labelled. In particular, each curve in the $(\alpha^*, \alpha)$-plane in Figure 3 corresponds to the locus of points which satisfy the equation $\mathcal{P}(> \alpha^*) - \mathcal{P}(> \alpha) = \text{constant}$, if $\alpha > \alpha^*$, or $\mathcal{P}(> \alpha^*) - \mathcal{P}(> \alpha) = \text{constant}$, if $\alpha < \alpha^*$, where the labels indicate the values of the constant.

## 4. Evaluation of $\sigma$

### 4.1. Filtered Density Fluctuation Spectrum

The mean value $< \rho_V >$ as well as the probability distribution $\mathcal{P}_{\text{obs}}(\rho_V)$ and the integrated probability $\mathcal{P}(> \rho_V)$ have been expressed in equations (24)– (29) in terms of the variance $\sigma^2$ in the fluctuation distribution (21). Now we must consider how to calculate $\sigma$.

From equation (22), we have $\sigma^2 = \langle \delta^2 \rangle$. But the variance $\sigma^2$ which is appropriate for our purpose here, is that which reflects the range of wavenumbers which might possibly contribute to the formation of gravitational condensations that are large enough to lead to "astronomer formation." Only wavenumbers corresponding to density fluctuations encompassing such sufficiently large masses should be allowed to contribute. This implies that the appropriate $\sigma$ for our purpose here is one calculated by filtering the underlying density field to eliminate the contribution from small wavelengths. This is accomplished by smoothing the density

Table 1.  THE PROBABILITY THAT $\rho_V / < \rho_V >$ IS BETWEEN TWO VALUES[1]

| $\alpha_1$ | $\alpha_2$ | $\mathcal{P}(> \alpha_1) - \mathcal{P}(> \alpha_2)$ | $\alpha_1$ | $\alpha_2$ | $\mathcal{P}(> \alpha_1) - \mathcal{P}(> \alpha_2)$ |
|---|---|---|---|---|---|
| 0.000 | 0.001 | 0.00355 | 1.5 | 1.6 | 0.01499 |
| 0.001 | 0.002 | 0.00317 | 1.6 | 1.7 | 0.01366 |
| 0.002 | 0.003 | 0.00299 | 1.7 | 1.8 | 0.01245 |
| 0.003 | 0.004 | 0.00287 | 1.8 | 1.9 | 0.01140 |
| 0.004 | 0.005 | 0.00277 | 1.9 | 2.0 | 0.01044 |
| 0.005 | 0.006 | 0.00269 | 2.0 | 2.1 | 0.00959 |
| 0.006 | 0.007 | 0.00262 | 2.1 | 2.2 | 0.00882 |
| 0.007 | 0.008 | 0.00256 | 2.2 | 2.3 | 0.00813 |
| 0.008 | 0.009 | 0.00251 | 2.3 | 2.4 | 0.00750 |
| 0.009 | 0.010 | 0.00247 | 2.4 | 2.5 | 0.00693 |
|  |  |  | 2.5 | 2.6 | 0.00642 |
| 0.00 | 0.01 | 0.02820 | 2.6 | 2.7 | 0.00585 |
| 0.01 | 0.02 | 0.02268 | 2.7 | 2.8 | 0.00561 |
| 0.02 | 0.03 | 0.02023 | 2.8 | 2.9 | 0.00512 |
| 0.03 | 0.04 | 0.01858 | 2.9 | 3.0 | 0.00477 |
| 0.04 | 0.05 | 0.01731 |  |  |  |
| 0.05 | 0.06 | 0.01629 | 0.0 | 0.5 | 0.50350 |
| 0.06 | 0.07 | 0.01543 | 0.5 | 1.0 | 0.18691 |
| 0.07 | 0.08 | 0.01468 | 1.0 | 1.5 | 0.10271 |
| 0.08 | 0.09 | 0.01404 | 1.5 | 2.0 | 0.06294 |
| 0.09 | 0.10 | 0.01346 | 2.0 | 2.5 | 0.04097 |
|  |  |  | 2.5 | 3.0 | 0.02777 |
| 0.0 | 0.1 | 0.18090 | 3.0 | 3.5 | 0.01937 |
| 0.1 | 0.2 | 0.11175 | 3.5 | 4.0 | 0.01382 |
| 0.2 | 0.3 | 0.08510 | 4.0 | 4.5 | 0.01004 |
| 0.3 | 0.4 | 0.06863 | 4.5 | 5.0 | 0.00740 |
| 0.4 | 0.5 | 0.05712 | 5.0 | 5.5 | 0.00553 |
| 0.5 | 0.6 | 0.04850 | 5.5 | 6.0 | 0.00418 |
| 0.6 | 0.7 | 0.04178 | 6.0 | 6.5 | 0.00318 |
| 0.7 | 0.8 | 0.03638 | 6.5 | 7.0 | 0.00245 |
| 0.8 | 0.9 | 0.03197 | 7.0 | 7.5 | 0.00189 |
| 0.9 | 1.0 | 0.02828 | 7.5 | 8.0 | 0.00148 |
| 1.0 | 1.1 | 0.02518 | 8.0 | 8.5 | 0.00115 |
| 1.1 | 1.2 | 0.02253 | 8.5 | 9.0 | 0.00092 |
| 1.2 | 1.3 | 0.02024 | 9.0 | 9.5 | 0.00072 |
| 1.3 | 1.4 | 0.01825 | 9.5 | 10.0 | 0.00058 |
| 1.4 | 1.5 | 0.01652 | 10.0 | $\infty$ | 0.00249 |

[1]For $s = 1$, where $\alpha \equiv \rho_V / < \rho_V >$

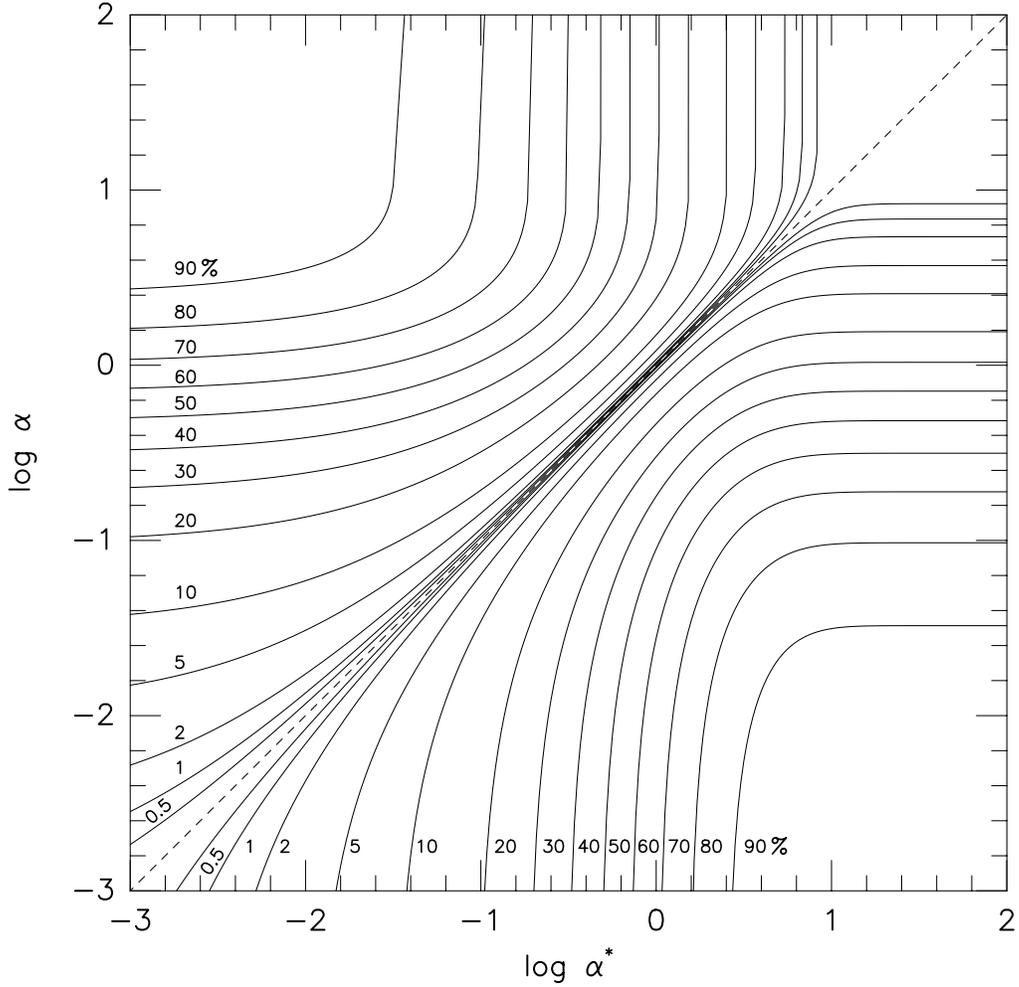

Fig. 3.— Interval Probability Isocontours. Each curve in the $(\alpha^*, \alpha)$-plane is the locus of points with a constant probability that a subuniverse is observed to have a value of $\rho_V / <\rho_V>$ in the range between $\alpha^*$ and $\alpha$ (i.e. $|\mathcal{P}(>\alpha^*) - \mathcal{P}(>\alpha)|$ = constant, where each curve is labelled with the value of this constant.)

field before calculating the variance $\sigma^2$, according to

$$\sigma^2 = \langle \tilde{\delta}^2(\mathbf{r}) \rangle \,, \tag{36}$$

where

$$\tilde{\delta}(\mathbf{r}) \equiv \int \delta(\mathbf{x}) W(\mathbf{x} - \mathbf{r}) d^3 x \,. \tag{37}$$

Here $W$ is a smoothing "window function," and $\mathbf{x}$ and $\mathbf{r}$ are co-moving coordinates, which following convention we shall normalize to give the proper distance at present. This yields the following familiar expression for the variance $\sigma^2$

$$\sigma^2 = \frac{1}{2\pi^2} \int_0^\infty P(k)\hat{W}^2(kR)k^2\,dk \,, \tag{38}$$

where $P(k)$ is the power spectrum (assuming statistical translation and rotation invariance)

$$P(|\mathbf{k}|) \equiv \int d^3x\,\langle\delta(\mathbf{x}+\mathbf{r})\delta(\mathbf{r})\rangle\,e^{i\mathbf{k}\cdot\mathbf{x}} \tag{39}$$

and $\hat{W}(kR)$ is the Fourier transform of the window function

$$\hat{W}(kR) \equiv \int d^3x\,e^{-i\mathbf{k}\cdot\mathbf{x}}W(x) \,, \tag{40}$$

with $R$ a length parameter to be specified below, introduced to make the argument of $\hat{W}$ dimensionless. (The bracket in eq. [39] implies an average over space; the assumptions of isotropy and homogeneity ensure that $P$ depends only on $k = |\mathbf{k}|$.) The window functions in which we are interested here are those which filter out modes of wavelength smaller than some characteristic value $R$. There are two conventional choices for the window function, the Gaussian window function,

$$\hat{W}_{\rm G}(u) = e^{-u^2/2} \,, \tag{41}$$

and the Top-Hat window function,

$$\hat{W}_{\rm TH}(u) = \frac{3}{u^3}(\sin u - u\cos u) \,, \tag{42}$$

(Peebles 1980). The baryonic mass associated with a density fluctuation of wavelength close to the filter scale is given by

$$M_f = \begin{cases} (2\pi)^{3/2}\rho_{B0}R^3 & \text{(Gaussian)}, \\ (4\pi/3)\rho_{B0}R^3 & \text{(Top-Hat)}, \end{cases} \tag{43}$$

with both $R$ and the cosmic mean baryon density $\rho_{B0}$ evaluated at the present. The radii for which both window functions enclose the same mass are thus related by

$$\frac{R_{\rm G}}{R_{\rm TH}} = \frac{(4\pi/3)^{1/3}}{(2\pi)^{1/2}} = 0.6431 \,. \tag{44}$$

We shall occasionally refer to a maximum wavenumber $k_{\max}$ as the wavenumber corresponding to a wavelength $R_{\rm G}$,

$$k_{\max} \equiv \frac{2\pi}{R_{\rm G}} = \frac{(2\pi)^{3/2}}{(4\pi/3)^{1/3}R_{\rm TH}} \,. \tag{45}$$

The particular value of $R_G$ (or $R_{TH}$) appropriate for use in calculating the mass fraction which collapses out of the background is uncertain. Our understanding of the detailed conditions necessary for the formation of planets and intelligent life has not yet advanced to the point of determining what the minimum mass condensation is which is capable of forming astronomers. Roughly speaking, we should filter out condensations that are too small to retain metals produced in the first generation of stars. The minimum mass condensation which is capable of this is currently unknown, however. It is not yet established, for example, whether globular clusters of mass $10^5 - 10^6 M_\odot$ are capable of self-enrichment, whereby a first generation of stars generates and releases heavy elements without expelling them from the cluster, so that they can subsequently be incorporated in a second generation of stars. Dwarf galaxies of even greater mass, in fact, are often postulated to undergo an initial burst of massive star formation which leads to supernova-driven expulsion of their interstellar gas (containing heavy elements). Even the typical galaxy in a rich cluster of galaxies is widely believed to have released most of its heavy elements into the intracluster medium, in order to account for the nearly solar metallicity of that gas, which dominates the baryonic mass of the cluster. In short, we do not currently know what the minimum mass scale (or associated wavelength of density fluctuations) is which satisfies the necessary condition that the metals produced by the first generation of stars are retained. Nor do we know if this is a *sufficient* condition for the formation of astronomers.

In fact, all we can say with certainty is that our own Milky Way galaxy met the necessary and sufficient conditions for forming planets, life, and astronomers. The Milky Way has a luminosity which makes it roughly an $L^*$-galaxy, the characteristic luminosity in the bright end of the galaxy luminosity function. If the minimum mass scale $M_f$ that can be responsible for astronomer formation corresponds to that of an $L^*$-galaxy, then the data on the galaxy luminosity function and the mass-to-light ratio of the bright inner parts of field spiral galaxies yields an estimate of the baryonic mass of the bright inner part of an $L^*$-galaxy of $M_f \approx 10^{11} h^{-1} M_\odot$ (see, for instance, Peebles 1993, pp. 122–123). This leads to an estimate of $R_G \cong h^{-1/3} (\Omega_B h^2 / 0.015)^{-1/3}$Mpc in present units, assuming a cosmic mean baryon density which is consistent with the current big bang nucleosynthesis abundance constraints. If, on the other hand, we take $M_f$ to be the mass of all the baryons initially within a comoving sphere whose volume equals that which, on average, typically contains just one $L^*$-galaxy today, this gives $R_G \cong 2h^{-1}$ Mpc. In view of the fact that the Milky Way is actually not an isolated galaxy, but is, instead, a member of the Local Group of galaxies, which includes more than one $L^*$-galaxy, we might even wish to consider the possibility that galaxy group membership is somehow essential to the formation of astronomers.[14] In that case, a value as

---

[14]For example, group membership might enable a galaxy which undergoes an early burst of star formation

large as $R_G \sim 3h^{-1}$ Mpc would even be reasonable.

In what follows, we will consider a range of values of $R_G$, therefore. For the case where $M_f$ corresponds to the bright inner part of an $L^*$-galaxy, we shall take $R_G = 1$ Mpc in present units (where, for simplicity, we shall drop the weak dependence of $R_G$ on $h$ for a fixed value of $\Omega_B h^2$). We will also bracket the range of possible outcomes by taking values of $R_G$ which are smaller and larger than this, respectively. On the low side, we take $R_G = 0.01$ Mpc, relevant for instance, if $M_f$ corresponds to the mass of a globular cluster. On the high side, we take $R_G = 2$ Mpc or 3 Mpc to illustrate the possibilities that $M_f$ corresponds either to the total mass within the mean volume per $L^*$-galaxy or else the mass of a small group of galaxies, respectively.

## 4.2.   The Cold Dark Matter Model

In order to evaluate $\sigma$ for a given value of $R_G$, we must adopt a model for the density fluctuation power spectrum at recombination. The cosmic microwave background anisotropy measured at large angles by the COBE satellite is consistent with Gaussian random noise density fluctuations with a scale-invariant primordial power spectrum $P(k) \propto k^n$, where $n = 1$, the case referred to as the Harrison–Zel'dovich spectrum. The range currently allowed by a statistical analysis of the first four years of data from the COBE DMR experiment is, in fact, $n = 1.2 \pm 0.3$ (Bennett et al. 1996). In what follows, we shall generally assume $n = 1$, which is the standard prediction of inflationary cosmology. Later, we can consider the effect of a "tilt" in the primordial spectrum away from the shape for $n = 1$. (Values of $n < 1$ can result, for example, if the primordial fluctuations include a gravitational wave contribution.)

In general, the power spectrum at recombination differs from the primordial shape $k^n$, except in the long wavelength limit measured directly by COBE. The difference reflects the linear growth of the density fluctuations prior to the recombination epoch, which is different for different wavelengths. The best-studied and most successful model for the growth of density fluctuations to date is the Cold Dark Matter (CDM) model. This model treats the CDM density fluctuations as adiabatic fluctuations in a cold, pressure-free gas. Since we are interested in the growth of density fluctuations in the baryon-electron fluid, which must be present to form stars, planets, and people, we make the assumption that this component collapses out in lock-step with the dark matter component, at least for density

---

to expel its heavy elements into the surrounding intra-group environment. The latter might then act as a reservoir from which the galaxy could later accrete some of its lost metals, after the expelled gas has cooled off.

fluctuations which are of wavelength large enough to behave in a pressure-free manner. As long as we restrict our attention to the epoch of recombination and later epochs and to wavelengths larger than the baryon Jeans length in the intergalactic medium, that is, the CDM and baryon power spectra should be identical. [For a detailed discussion of the effects of Jeans-mass- filtering on the linear growth of baryon density fluctuations in a flat, matter-dominated CDM model in which the Jeans mass is increased by the reheating of the intergalactic medium which accompanies its reionization, the reader is referred to Shapiro, Giroux, & Babul (1994)].

### 4.2.1. The Power Spectrum

We use for the CDM power spectrum the expression given by Liddle et al. (1996) and references therein:

$$P(k, z) = 2\pi^2 \left(\frac{c}{H_0}\right)^{3+n} (\delta_H)^2 \, k^n \, T^2(q) A^{-2}(z, 0) \,, \tag{46}$$

where

$$A(z, 0) = \frac{\delta_+(0)}{\delta_+(z)} \,, \tag{47}$$

$$T(q) = \frac{\ln(1 + 2.34q)}{2.34q} \Big[1 + 3.89q + (16.1q)^2 + (5.46q)^3 + (6.71q)^4\Big]^{-1/4} \,, \tag{48}$$

$$q = \frac{k}{h\Gamma \text{ Mpc}^{-1}} \,, \tag{49}$$

$$\Gamma = \Omega_0 h e^{-\Omega_{B0} - \Omega_{B0}/\Omega_0} \,, \tag{50}$$

and $\delta_H$ is the dimensionless amplitude at horizon crossing, which must be taken from observations of anisotropies in the microwave radiation background; $\Omega_0$ is the total matter density parameter ($\Omega_0 = \rho_0/\rho_{\text{crit},0}$, where $\rho_{\text{crit},0} = 3H_0^2/8\pi G$); $\Omega_{B0}$ is the corresponding parameter for baryonic matter; $H_0$ is the Hubble constant; $h = H_0/100 \, \text{km s}^{-1} \text{Mpc}^{-1}$; $\delta_+$ is the pure growing mode solution for the evolution of linear density fluctuations in this flat universe with nonzero cosmological constant; and $A(z, 0) = \delta_+(0)/\delta_+(z)$ is the linear growth factor between redshift $z$ and the present. For $n = 1$, these formulae describe the case of the Harrison–Zel'dovich scale-invariant power- law primordial spectrum, modified by the growth of fluctuations in a CDM model universe, for a flat universe with a nonzero cosmological constant $\lambda_0 = 1 - \Omega_0 = \rho_V/\rho_{\mathrm{crit},0}$. The fitting formula for $T(q)$ is from Bardeen et al. (1986), but with $\Gamma$ given by a fit by Sugiyama (1995) in the form quoted by Liddle et al. (1996). (The numerical coefficients in this formula depend on the present microwave radiation energy energy density and on the quantities $100$ km/sec and $1$ Mpc used in defining the dimensionless quantities $q$ and $h$, but not on $\rho_0$ or $H_0$.) The formula for $\Gamma$ given in equation (50) includes an exponential correction factor for the effect of nonzero baryon density. Since the variable $\Gamma$ is often used in the literature to refer, instead, just to the product $\Omega_0 h$ (i.e. without the exponential correction factor), the so-called "shape parameter" for CDM models, we will also define $\Gamma_0 \equiv \Omega_0 h$, to use whenever we wish to refer only to this product. In the following calculations, we use $\Omega_{B0} = 0.015 h^{-2}$, consistent with big bang nucleosynthesis constraints from the abundance of light elements (e.g. Copi, Schramm, & Turner 1995).

It is important to note that the explicit dependences of equations (46)–(50) on the values of $\Omega_0$ and $h$ do not mean that the power spectrum at recombination is different for different subuniverses with different values of the cosmological constant. All factors which depend upon $\Omega_0$ and $h$ reflect the fact that a knowledge of the local values of $\Omega_0$ and $h$ in our *own* subuniverse is required in order to *interpret* present-day observations in our own subuniverse (such as those of cosmic microwave background anisotropy) unambiguously to determine the power spectrum at recombination *assumed common to all subuniverses*. We must, therefore, distinguish clearly between the particular values of $\Omega_0 = 1 - \lambda_0$ and $h$ in our own subuniverse, on which our inference of the universal power spectrum depends, and the variables $\Omega_0 = 1 - \lambda_0$ and $h$, different for different subuniverses, on which the probability of galaxy formation in any subuniverse depends. To avoid any possible confusion on this point, we will, henceforth, indicate the values of these quantities in our *own* subuniverse by adding an asterisk to the symbol (i.e. $\lambda_0^*$, $\Omega_0^*$, $H_0^*$, $\rho_V^*$, etc.). This notation is not necessary for the quantities $\bar{\rho}$, $\bar{\rho}_B$, $P(k)$, or $\sigma$, however, since these are assumed not to vary from one subuniverse to another.

### 4.2.2. *The Linear Growth Rate in a Flat Universe with Nonzero Cosmological Constant*

In order to evaluate equation (46) for the CDM power spectrum for any particular value of the vacuum energy density, we must evaluate the growth factor $A(z,0;\lambda_0)$ in equation (47) as a function of $z$ and $\lambda_0 \equiv \rho_V/\rho_{\mathrm{crit},0}$. It is convenient to express this in term of a function $f(\lambda_0, z)$, defined as the ratio of the growth factor $(1+z)$ in an Einstein–de Sitter universe to $A(z,0;\lambda_0)$:

$$f(\lambda_0, z) \equiv \frac{1+z}{A(z,0)} = (1+z)\frac{\delta_+(z)}{\delta_+(0)}\,, \tag{51}$$

where $\delta_+$ is the amplitude of the linear growing mode, which is given for general $\lambda_0 \neq 0$ by

$$\delta_+(z; \lambda_0) = \left(\frac{1}{y}+1\right)^{1/2} \int_0^y \frac{dw}{w^{1/6}(1+w)^{3/2}} \tag{52}$$

(Martel 1991), with

$$y \equiv \frac{\rho_V}{\bar{\rho}(z)} = \frac{\lambda_0}{\Omega_0}(1+z)^{-3}\,, \tag{53}$$

and $\Omega_0 \equiv 1 - \lambda_0$. Using equations (51)–(53), we get, after some algebra,

$$
\begin{aligned}
f(\lambda_0, z) &= \Omega_0^{1/2}(1+z)^{5/2}\left[1 + \frac{\lambda_0}{\Omega_0(1+z)^3}\right]^{1/2}\left[\int_0^{\lambda_0/\Omega_0}\frac{dw}{w^{1/6}(1+w)^{3/2}}\right]^{-1} \\
&\times \int_0^{\frac{\lambda_0}{\Omega_0(1+z)^3}}\frac{dw}{w^{1/6}(1+w)^{3/2}}\,.
\end{aligned} \tag{54}
$$

For $1 + z \gg 1$, this gives the $z$-independent result

$$f(\lambda_0, z) \simeq f(\lambda_0) = \frac{6\lambda_0^{5/6}}{5\Omega_0^{1/3}}\left[\int_0^{\lambda_0/\Omega_0}\frac{dw}{w^{1/6}(1+w)^{3/2}}\right]^{-1}\,. \tag{55}$$

The corrections are of order $(1+z)^{-3}$, which for the case $z \approx 1000$ that interests us is entirely negligible. We have evaluated the integral in equation (55) numerically, and have plotted the function $f(\lambda_0)$ in Figure 4. As we see, $f(\lambda_0)$ differs substantially from unity only for relatively large values of $\lambda_0$.

### 4.2.3. Normalization at Recombination from Cosmic Microwave Background Anisotropy Measurements

According to Bunn & White (1996), the first four years of data on the cosmic microwave background temperature anisotropy detected by the COBE DMR experiment may be fit with a dimensionless amplitude at horizon crossing given by the formula

$$\delta_H^* = 1.94 \times 10^{-5}(\Omega_0^*)^{-0.785 - 0.05\ln\Omega_0^*}\exp\left[a(n-1) + b(n-1)^2\right]\,, \tag{56}$$

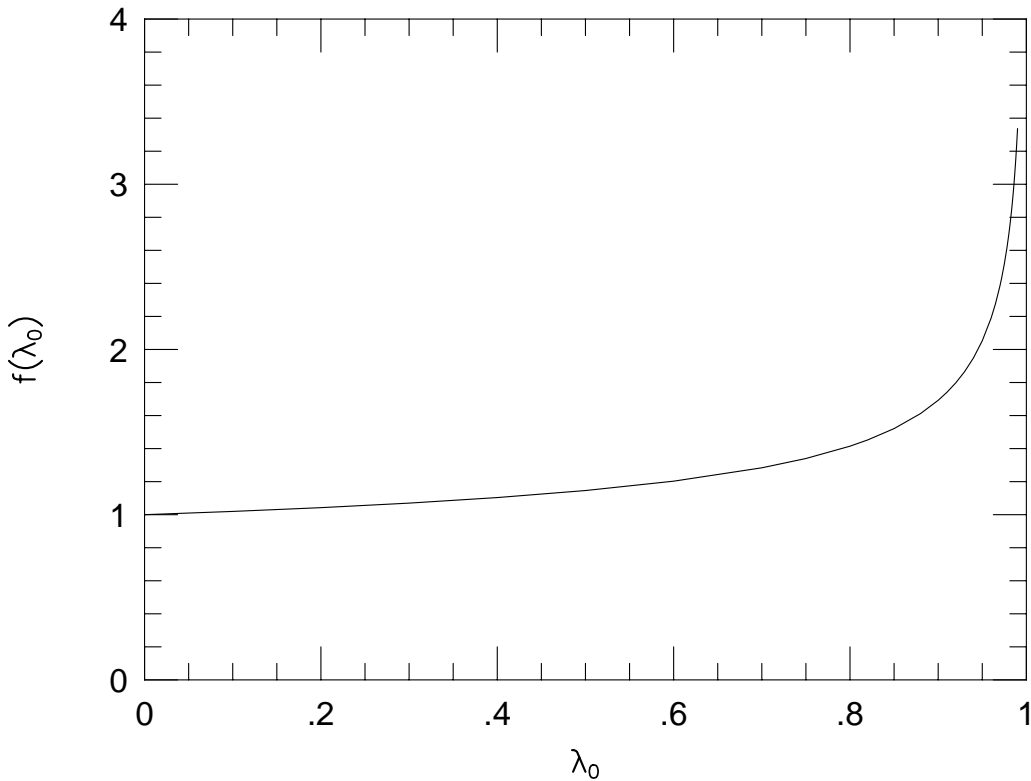

Fig. 4.— Ratio $f(\lambda_0)$ of the linear growth factors for the Einstein-de Sitter model and the flat $\lambda_0 \neq 0$ model defined by equation (55), versus $\lambda_0$.

where an asterisk, recall, denotes that the quantities are evaluated for our own subuniverse only. There are two sets of values of the constants $a$ and $b$, which correspond to the cases of $n \neq 1$ without any gravitational wave contribution ($a = -0.95$, $b = -0.169$) and of power-law inflation with gravitational waves ($a = 1$, $b = 1.97$), respectively.

Equations (46)–(51), along with equations (55) and (56), can now be evaluated to compute the power spectrum at recombination for any flat model for any values of $\lambda_0^*$ (or, equivalently, $\Omega_0^*$) and $h^*$. The results for $n = 1$ are shown in Figure 5 for $\lambda_0^* = 0$ and $h^* = 0.5$ or 1 (top panel), and for various values of $\lambda_0^*$ and $h^* = 0.5$ (bottom panel).

### 4.3. Results for $\sigma$ and $\sigma^3 \bar{\rho}$

The variance $\sigma^2$ at recombination is given by equations (38) and (46)–(51), as

$$
\begin{aligned}
\sigma(z_{\mathrm{rec}}) &= (c_{100})^{(n+3)/2} \Big[ \Gamma^{(n+3)/2} \delta_H A(z_{\mathrm{rec}}, 0)^{-1} K_n^{1/2}(q_{\max}) \Big]_* \\
&= (c_{100})^{(n+3)/2} (1 + z_{\mathrm{rec}})^{-1} \Big[ \Gamma^{(n+3)/2} \delta_H f(\lambda_0) K_n^{1/2}(q_{\max}) \Big]_* ,
\end{aligned}
\tag{57}
$$

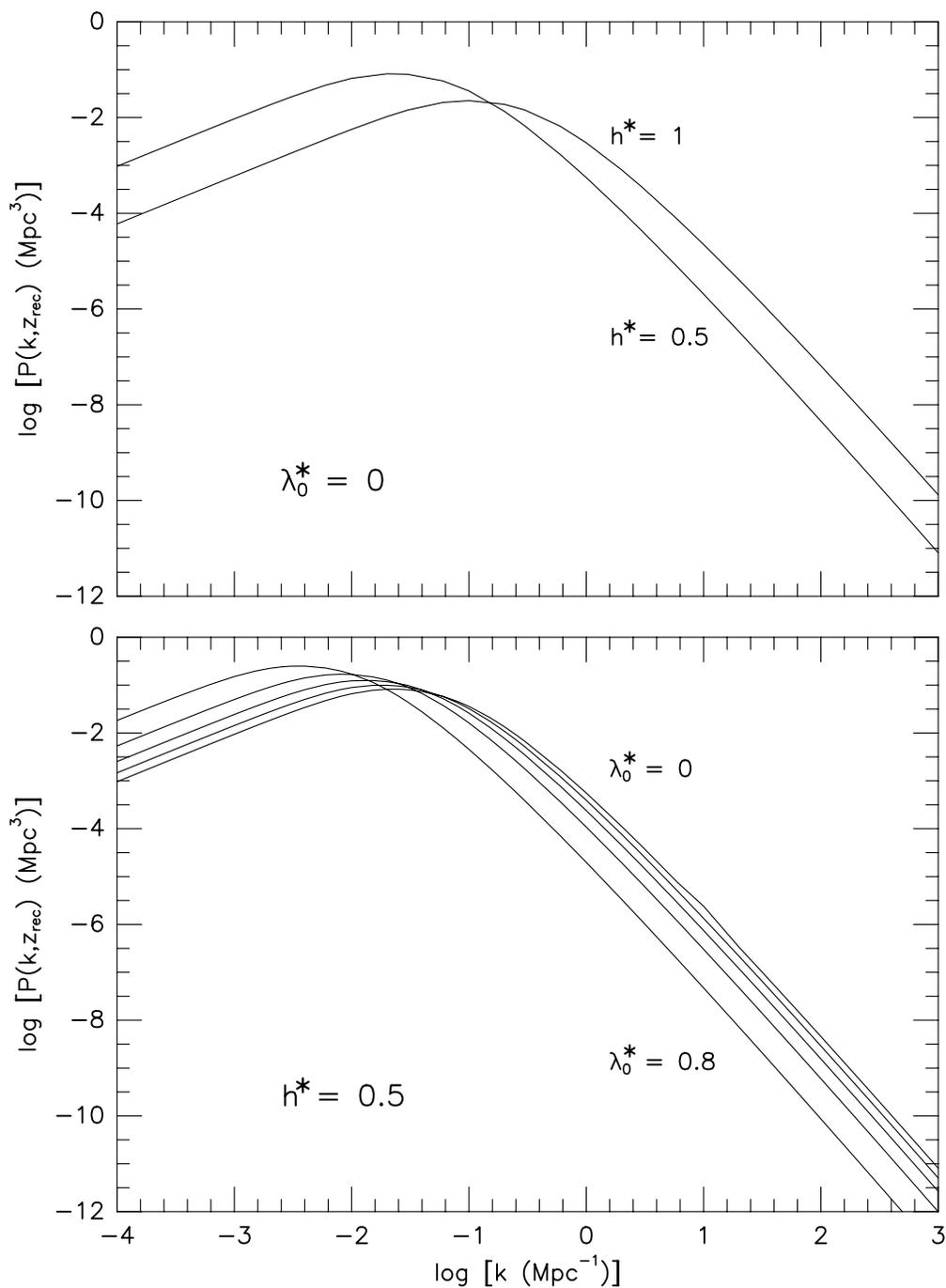

Fig. 5.— CDM power spectrum at recombination ($z_{\rm rec} = 1000$), for $n = 1$, versus comoving wavenumber $k$ ($= 2\pi/\lambda$, where wavelength $\lambda$ is in present units of Mpc) . Top panel: Einstein-de Sitter model ($\Omega_0^* = 1$, $\lambda_0^* = 0$) with Hubble constants $h^* = 0.5$ and 1. Bottom panel: Flat models with $h^* = 0.5$ and $\lambda_0^* = 0$, 0.2, 0.4, 0.6, and 0.8.

where the symbol "$*$" labelling the brackets in equation (57) and in what follows, indicates that all quantities inside the brackets are evaluated using the values $\lambda_0^*$ and $h^*$ in our own subuniverse; $c_{100} = 2997.9$ is the speed of light in units of 100 km/sec; the second equality refers to the result to leading order in $(1 + z_{rec})^{-3}$; and

$$K_n(q_{max}) \equiv \begin{cases} \int_0^\infty q^{n+2} T^2(q) \hat{W}_G^2 \left( 2\pi \dfrac{q}{q_{max}} \right) dq & \text{(Gaussian)} \\ \int_0^\infty q^{n+2} T^2(q) \hat{W}_{TH}^2 \left[ \dfrac{(2\pi)^{3/2}}{(4\pi/3)^{1/3}} \dfrac{q}{q_{max}} \right] dq & \text{(Top-Hat)} \end{cases} \quad (58)$$

The integrals in equation (58) are evaluated numerically. The results are shown in Figure 6 including the values $n = 1$, 0.9, and 0.8. (Note: For $n \neq 1$, we hereafter adopt constants $a$ and $b$ for the case of $n \neq 1$ with no gravitational waves. The case *with* gravitational waves yields a slightly smaller value of $\sigma$ for the same value of $n$.)

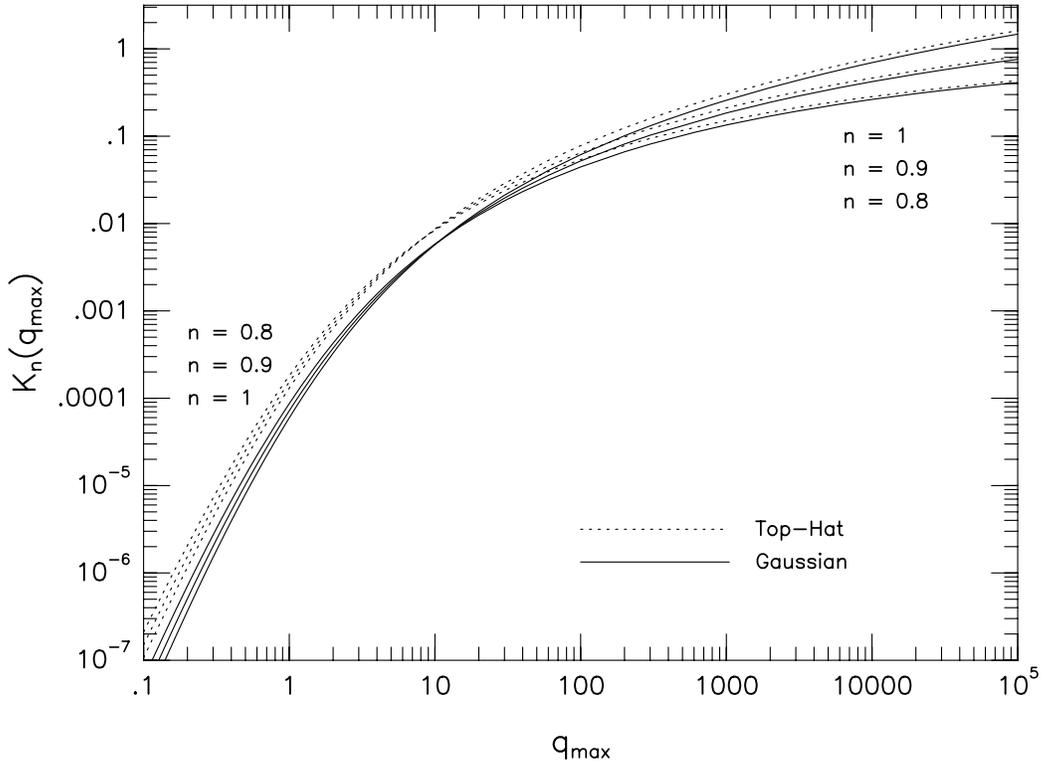

Fig. 6.— The dimensionless integral $K_n(q_{max})$ vs. $q_{max}$, defined by equation (58) for CDM density fluctuations, for the Gaussian (solid) and Top-Hat (dashed) window functions, respectively, for $n = 1$, 0.9, and 0.8, as labelled.

It is customary to report the normalizations of the power spectra for different models in terms of the value of $\sigma$ evaluated at the present for a particular filter scale, assuming fluctuations continue to grow at the linear growth rate. For our purpose here, however, we must evaluate $\sigma$ at recombination, thereby undoing the effects of the growth of fluctuations since that epoch which influence the value of $\sigma$ at the present. Our goal, recall, is to use the observations of the cosmic microwave background anisotropy made by astronomers in our *own* subuniverse to infer the universal density fluctuation distribution, common to all subuniverses at $z_{rec}$. Unfortunately, our ability to infer this universal density fluctuation distribution is limited by the fact that we must know the values of $\lambda_0^*$ and $h^*$ in our *own* subuniverse in order to interpret the cosmic microwave background anisotropy measurements unambiguously. We illustrate the dependence of the inferred density fluctuations on the assumed values of $\lambda_0^*$ and $h^*$ in Figure 7, where we have plotted the value of $\sigma$ at recombination as a function of $\lambda_0^*$, for $R_G = 0.01, 1, 2,$ and $3\,\mathrm{Mpc}$, for $n = 1$, 0.9, and 0.8. We have taken $z_{rec} = 1000$ for the results in Figure 7. The effect of a "tilt" to $n < 1$ is to decrease $\sigma$ relative to its value for $n = 1$, for the same value of $q_{max}$, or equivalently, of $R_G$.

As already mentioned, results for the probability distribution of $\rho_V$ actually depend not on $\sigma$ or $\bar{\rho}$, but on the parameter $\sigma^3\bar{\rho}$. The total matter density $\bar{\rho}$ at recombination is related to the present matter density $\rho_0$ by $\bar{\rho} = \rho_0(1 + z_{rec})^3$, so as promised $\sigma^3\bar{\rho}$ is independent of the precise value chosen for $z_{rec}$:

$$\sigma^3\bar{\rho} = \rho_0^*(c_{100})^{(3n+9)/2}\left[\Gamma^{(n+3)/2}\delta_H f(\lambda_0)K_n^{1/2}(q_{max})\right]_*^3. \tag{59}$$

## 5. RESULTS

With all of the ingredients necessary to evaluate the probability distribution $\mathcal{P}_{obs}(\rho_V)$ thus assembled, we can now evaluate the probability of observing any particular value of $\rho_V$ anywhere in the universe, as well as the average and median values observed, as functions of the values we adopt for $\rho_V^*$ (or $\lambda_0^* = 1 - \Omega_0^*$) and $h^*$ in our own subuniverse. There are two ways to use this information. We may try to guess the actual value of $\rho_V^*$, by *assuming* that we live in a typical subuniverse, in which $\rho_V^*$ is equal to either the mean or the median observed values of $\rho_V$ for all astronomers in all subuniverses. (This will be done in subsections 5.1 and 5.2.) Such a "prediction" carries low confidence, since it is always possible that $\rho_V^*$ in our subuniverse could be significantly different from the anthropic mean or median. Alternatively, we can use what data we have to estimate the range of observationally allowed values of $\rho_V^*$, and then calculate the likelihood that a randomly chosen astronomer in any subuniverse would find such a value. (This will be the subject of subsections 5.3 and 5.4.) If this likelihood proves to be appreciable, then the anthropic principle would survive as a possible explanation for the particular value of the cosmological constant in our own

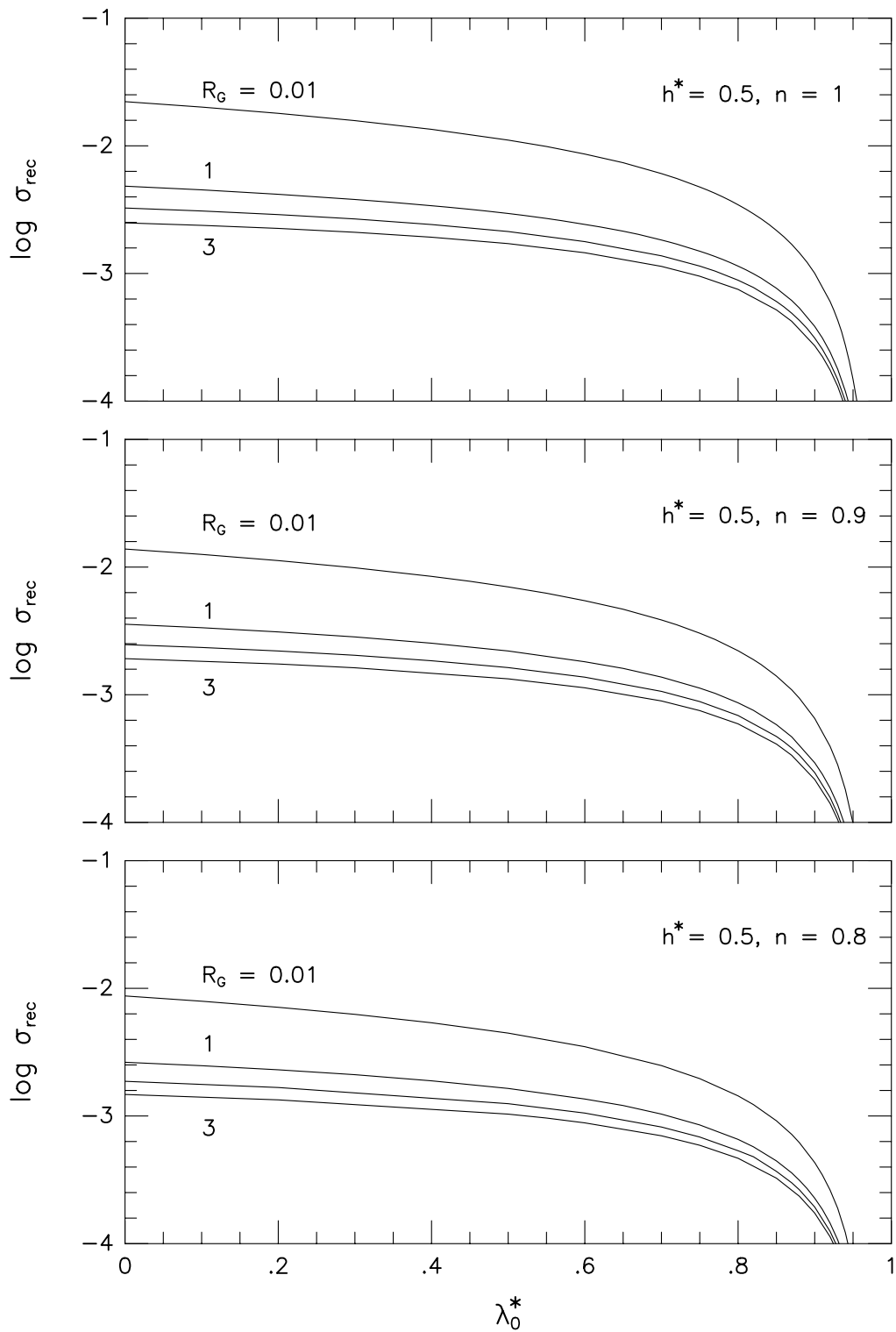

Fig. 7.— (a) The rms density fluctuation at recombination (i.e. at $z_{rec} = 1000$), $\sigma = \sigma_{rec}$, in the COBE-normalized, flat CDM model, versus $\lambda_0^*$, for $R_G = 0.01, 1, 2,$ and $3\,\mathrm{Mpc}$, respectively, as labelled, for $h^* = 0.5$, for $n = 1$ (top panel), 0.9 (middle panel), and 0.8 (bottom panel). (b) Same as Fig. 7(a), except for $h^* = 1$.

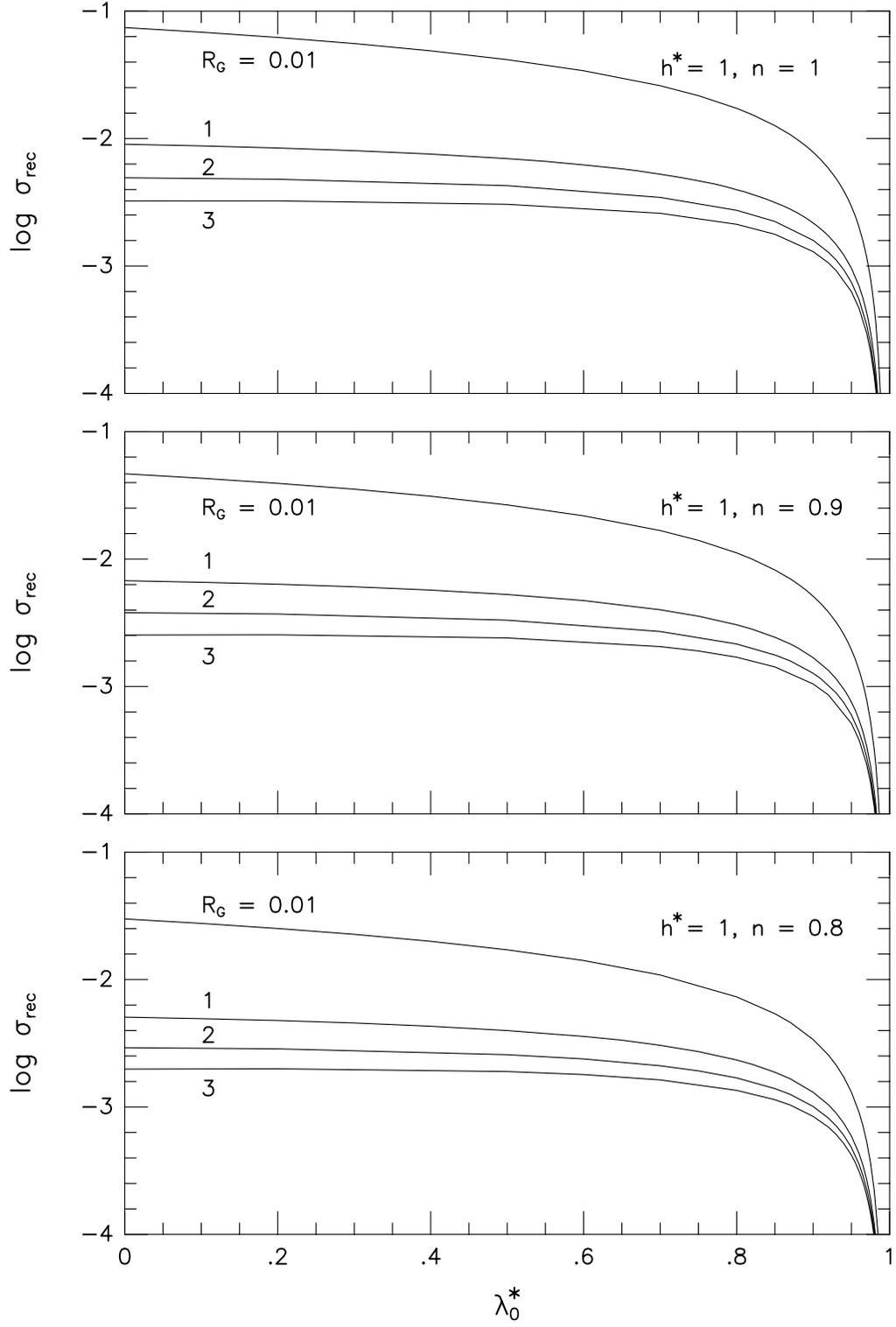

Fig. 7b

subuniverse.

## 5.1. The Average Observed Value of the Vacuum Energy Density in a COBE-Normalized CDM Universe

From equations (24) and (59), the average observed vacuum density is given by

$$\frac{<\rho_V>}{\rho_0^*} = \left[\frac{625(2\pi)^{1/2}}{243}\right] \frac{I_1(s)}{I_0(s)} c_{100}^{(3n+9)/2} \left\{ \Gamma^{(3n+9)/2} \left[f(\lambda_0)\delta_H\right]^3 K_n^{3/2}(q_{\max}) \right\}_* . \qquad (60)$$

where again the symbol $*$ indicates that all quantities inside the braces are evaluated at the particular values $\lambda_0 = \lambda_0^*$, $\Omega_0 = \Omega_0^*$, $h = h^*$ for our own subuniverse.

We have used equation (60) to plot the ratio $\rho_V^*/ <\rho_V>$ versus the assumed value $\lambda_0^*$ of the cosmological constant in our subuniverse, and also versus the ratio $\lambda_0^*/\Omega_0^*$, in Figure 8 for $h^* = 0.5$ and 1, $R_G = 0.01, 1, 2,$ and 3 Mpc, and $n = 1, 0.9, 0.8$.

*For what values of $\lambda_0^*$ and $h^*$ is the vacuum energy density in our own subuniverse equal to the observed average of all the subuniverses?* For the particular values of $R_G$ and $h^*$ assumed in plotting the curves in Figure 8, the intersection of the horizontal dashed line with the curves indicates the values for which $\rho_V^* = <\rho_V>$. To answer this question more generally as a function of $R_G$ and $h^*$, we have solved the implicit equation $\rho_V^* = <\rho_V>$ numerically by setting the quantity (60) equal to $\rho_V^*/\rho_0^* = \lambda_0^*/(1 - \lambda_0^*)$, using the secant method. The results are shown in Table 2 for $n = 1$ and $s = 1$. According to these results, if the probability that observers are created in any subuniverse is proportional to the amount of mass which eventually collapses out into the bright inner parts of $L^*$-galaxy-mass objects or larger objects (i.e. $R_G \sim 1$ Mpc), then our own subuniverse has the average observed value of the vacuum energy density if the value of $\lambda_0$ which we observe locally is

$$\frac{\rho_V^*}{\rho_0^*} \cong \begin{cases} 4.1 \ , \\ 7.3 \ , \\ 12.4 \ , \end{cases} \qquad \lambda_0^* \cong \begin{cases} 0.80 \ , & h^* = 0.5 \ ; \\ 0.88 \ , & h^* = 0.7 \ ; \\ 0.93 \ , & h^* = 1 \ . \end{cases} \qquad (61)$$

If the relevant collapsed fraction is, instead, that which condenses into objects as large as or larger than those which contain the mean total baryon mass per $L^*$-galaxy (i.e. $R_G \sim 2$ Mpc), then our own universe has $\rho_V^* = <\rho_V>$ if

$$\frac{\rho_V^*}{\rho_0^*} \cong \begin{cases} 3.4 \,, \\ 5.9 \,, \\ 9.8 \,, \end{cases} \qquad \lambda_0^* \cong \begin{cases} 0.77 \,, & h^* = 0.5 \,; \\ 0.86 \,, & h^* = 0.7 \,; \\ 0.92 \,, & h^* = 1 \,. \end{cases} \qquad (62)$$

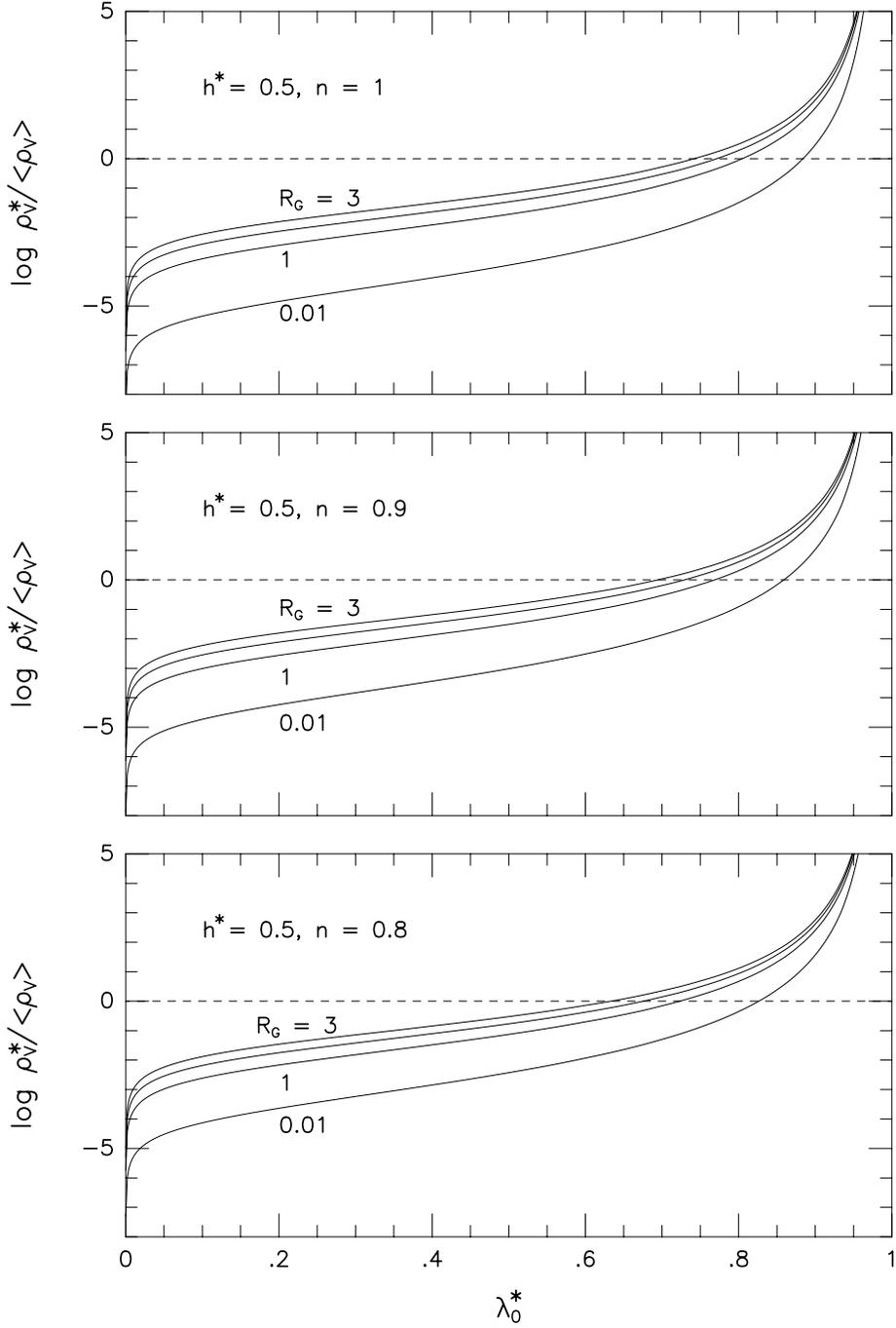

Fig. 8.— (a) The vacuum energy density adopted for our own subuniverse, $\rho_V^* = 3(H_0^*)^2\lambda_0^*/8\pi G$, is plotted in units of the mean value of the vacuum energy density observed in all subuniverses, $<\rho_V>$, versus $\lambda_0^*$, the cosmological constant in our own subuniverse, for local Hubble constant $h^* = 0.5$, for a COBE-normalized flat CDM model with primordial power spectrum index $n = 1$ (top panel), 0.9 (middle panel), and 0.8 (bottom panel), assuming shape parameter $s = 1$. Intersections of the dashed horizontal lines at $\rho_V^*/<\rho_V> = 1$ with the various curves give the solution of the implicit equation obtained by setting $<\rho_V> = \rho_V^*$ in equation (60); (b) Same as Fig. 8(a), except plotted versus $\lambda_0^*/\Omega_0^*$; (c) Same as Fig. 8(a), except for $h^* = 1$; (d) Same as Fig. 8(b), except for $h^* = 1$.

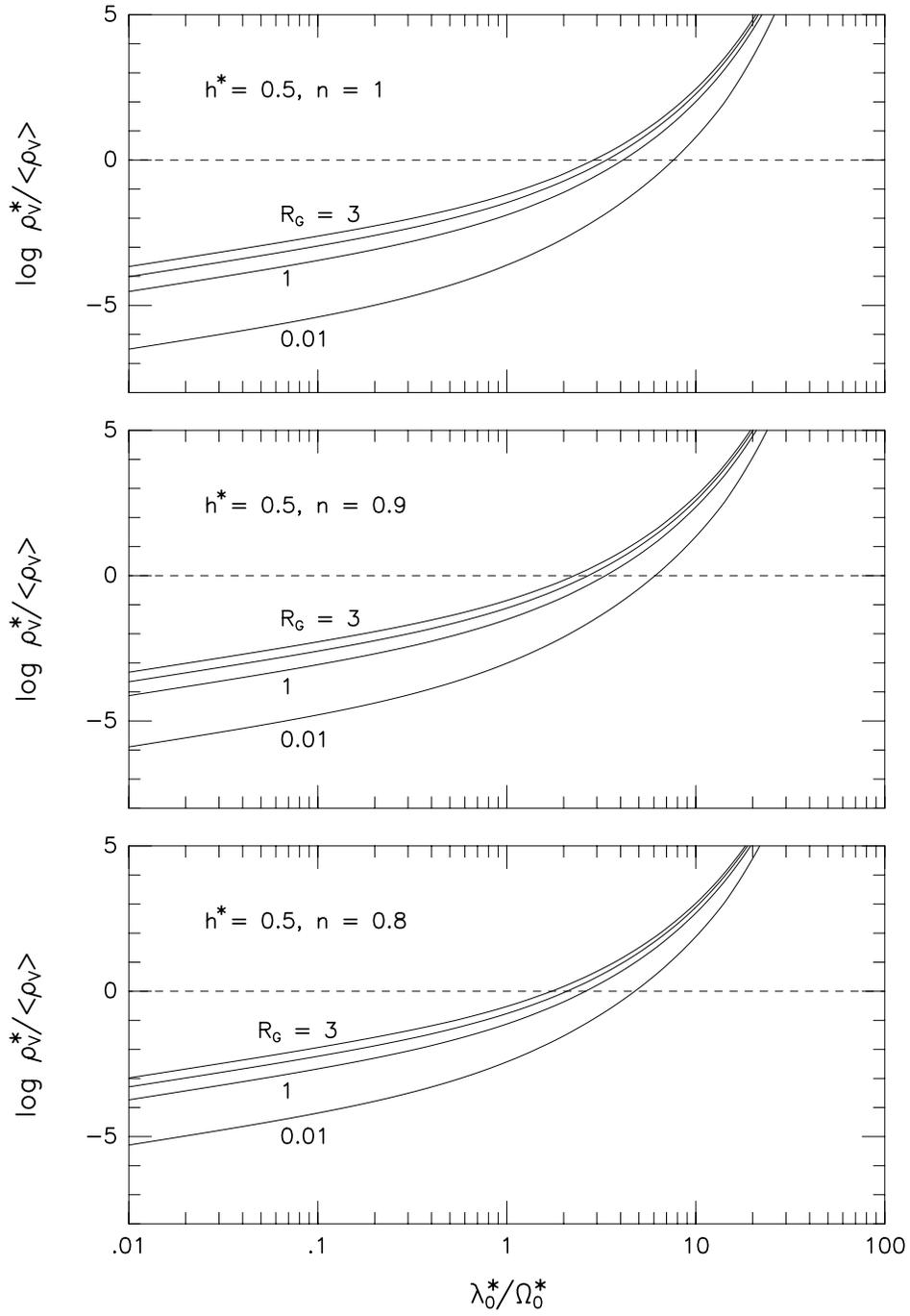

Fig. 8b

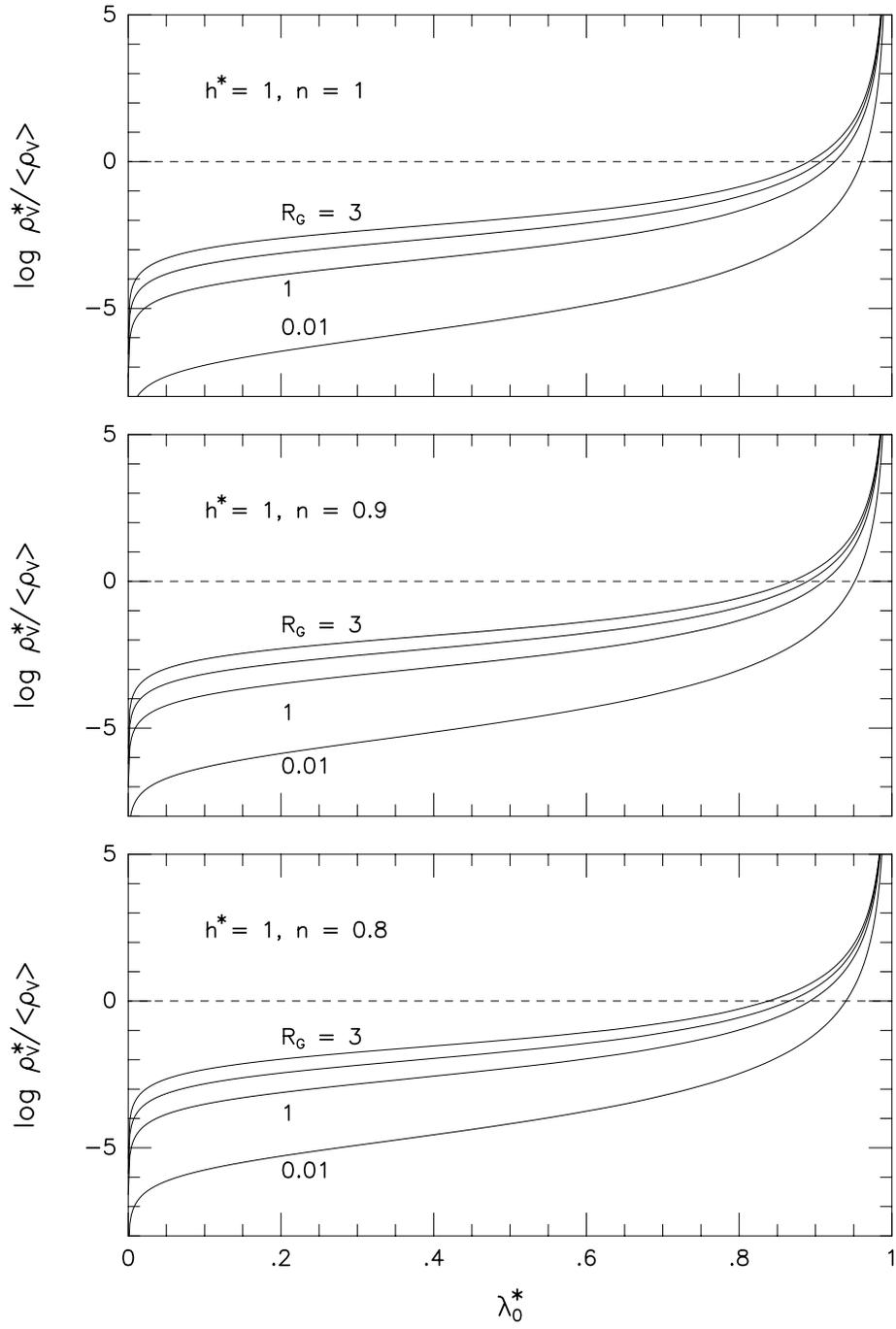

Fig. 8c

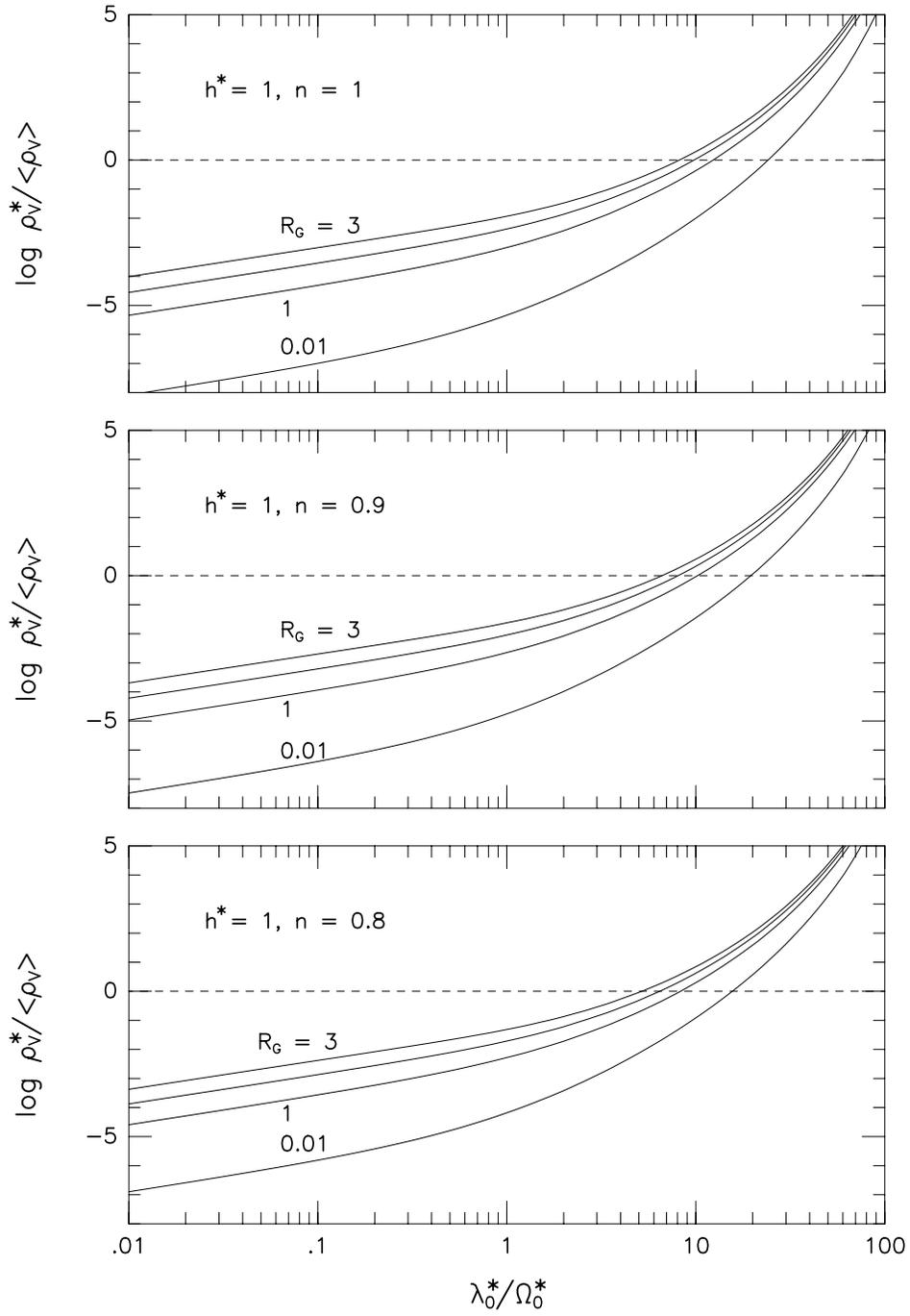

Fig. 8d

Table 2.  LOCAL VACUUM ENERGY DENSITIES WHICH EQUAL THE GLOBAL
AVERAGE $(\rho_V^* = <\rho_V>)$[1]

| $R_{\rm G}$ (Mpc)[2] | $h^* = 0.5$ | | | $h^* = 0.7$ | | | $h^* = 1$ | | |
|---|---|---|---|---|---|---|---|---|---|
| | $\lambda_0^*$ | $\Omega_0^*$ | $\lambda_0^*/\Omega_0^*$ | $\lambda_0^*$ | $\Omega_0^*$ | $\lambda_0^*/\Omega_0^*$ | $\lambda_0^*$ | $\Omega_0^*$ | $\lambda_0^*/\Omega_0^*$ |
| 0.003 | 0.8905 | 0.1095 | 8.1324 | 0.9373 | 0.0627 | 14.949 | 0.9635 | 0.0365 | 26.397 |
| 0.010 | 0.8839 | 0.1161 | 7.6133 | 0.9327 | 0.0673 | 13.859 | 0.9602 | 0.0398 | 24.126 |
| 0.030 | 0.8740 | 0.1260 | 6.9365 | 0.9256 | 0.0744 | 12.441 | 0.9562 | 0.0438 | 21.831 |
| 0.100 | 0.8593 | 0.1407 | 6.0173 | 0.9161 | 0.0839 | 10.919 | 0.9500 | 0.0500 | 19.000 |
| 0.300 | 0.8397 | 0.1603 | 5.2383 | 0.9032 | 0.0968 | 9.3306 | 0.9415 | 0.0585 | 16.094 |
| 1.000 | 0.8046 | 0.1954 | 4.1177 | 0.8791 | 0.1209 | 7.2713 | 0.9252 | 0.0748 | 12.369 |
| 2.000 | 0.7712 | 0.2288 | 3.3706 | 0.8554 | 0.1446 | 5.9156 | 0.9073 | 0.0927 | 9.7875 |
| 3.000 | 0.7418 | 0.2582 | 2.8730 | 0.8339 | 0.1661 | 5.0205 | 0.8912 | 0.1088 | 8.1912 |
| 6.000 | 0.6627 | 0.3373 | 1.9647 | 0.7669 | 0.2331 | 3.2900 | 0.8326 | 0.1674 | 4.9737 |
| 10.000 | 0.5424 | 0.4576 | 1.1853 | 0.6408 | 0.3592 | 1.7840 | 0.6680 | 0.3320 | 2.0120 |
| 15.000 | 0.3582 | 0.6418 | 0.5581 | 0.3539 | 0.6461 | 0.5478 | 0.1665 | 0.8335 | 0.1998 |
| 20.000 | 0.1861 | 0.8139 | 0.2287 | 0.1217 | 0.8783 | 0.1386 | 0.0377 | 0.9623 | 0.0918 |

[1] For Harrison-Zel'dovich scale-invariant primordial power spectrum ($n = 1$), and $s = V/U = 1$.

[2] Present units

If, instead, galaxy groups are the minimum scale of interest (i.e. $R_G \sim 3\,\text{Mpc}$), then our own subuniverse has $\rho_V^* = <\rho_V>$ if

$$\frac{\rho_V^*}{\rho_0^*} \cong \begin{cases} 2.9 \,, \\ 5.0 \,, \\ 8.2 \,, \end{cases} \qquad \lambda_0^* \cong \begin{cases} 0.74 \,, & h^* = 0.5 \,; \\ 0.83 \,, & h^* = 0.7 \,; \\ 0.89 \,, & h^* = 1 \,. \end{cases} \qquad (63)$$

On the other hand, if globular cluster formation is enough to satisfy the anthropic constraint (i. e., $R_G \sim 0.01$ Mpc), then our own universe has $\rho_V^* = <\rho_V>$ if

$$\frac{\rho_V^*}{\rho_0^*} \cong \begin{cases} 7.6 \,, \\ 13.9 \,, \\ 24.1 \,, \end{cases} \qquad \lambda_0^* \cong \begin{cases} 0.88 \,, & h^* = 0.5 \,; \\ 0.93 \,, & h^* = 0.7 \,; \\ 0.96 \,, & h^* = 1 \,. \end{cases} \qquad (64)$$

We have assumed $s = 1$ in all these cases.

The effect of the "tilt" from $n = 1$ to $n < 1$ on these results is to reduce the values of $<\rho_V>$, or $<\lambda_0/\Omega_0>$, somewhat. This can be understood in terms of the ratio $\sigma_n^3/\sigma_1^3$, where the subscript refers to the index of the primordial power spectrum. This ratio is given by

$$\begin{aligned} \frac{\sigma_n^3}{\sigma_1^3} &= \left( \frac{c\Gamma^* h^*}{H_0^*} \text{Mpc}^{-1} \right)^{3(n-1)/2} \left[ \frac{\delta_H^*(n)}{\delta_H^*(1)} \right]^3 \left[ \frac{K_n(q_{max})}{K_1(q_{max})} \right]^{3/2} \\ &\cong (c_{100}\Gamma^*)^{3(n-1)/2} \exp\left[ 3a(n-1) + 3b(n-1)^2 \right] \left[ \frac{K_n(q_{max})}{K_1(q_{max})} \right]^{3/2} \,. \end{aligned} \qquad (65)$$

The exponential factor in the second line of equation (65) is always close to unity for $1-n \ll 1$. The first factor in this line, however, decreases as $(1-n)$ increases. For $q_{max} \gtrsim 10$, the results plotted in Figure 6 show that the ratio $K_n/K_1$ also decreases as $1-n$ increases, albeit slowly for $1-n \ll 1$. For $\Gamma^* = 0.2$, $h^* = 0.5$, and $R_G = 1\,\text{Mpc}$ (i.e. $q_{max} \cong 63$), for example, we find that $\sigma_n^3/\sigma_1^3 = 0.416$ and $0.173$ for $n = 0.9$ and $0.8$, respectively. A small tilt to $n < 1$, therefore, decreases the variance of the density fluctuations on galaxy-mass scales. For a tilt within the range allowed by the cosmic microwave background anisotropy measured at large angles by the COBE satellite, this effect of the tilt translates into a modest *decrease* in the average $\rho_V$ observed for all subuniverses. For example, our own subuniverse has the average observed value of $\rho_V$ (i.e. $\rho_V^* = <\rho_V>$) for $R_G = 1\,\mathrm{Mpc}$ and $h^* = 0.7$ if

$$\lambda_0^* \cong \begin{cases} 0.827, & n = 0.8; \\ 0.857, & n = 0.9; \\ 0.879, & n = 1. \end{cases} \tag{66}$$

If $R_G = 2\,\mathrm{Mpc}$ and $h^* = 0.7$, this becomes

$$\lambda_0^* \cong \begin{cases} 0.792, & n = 0.8; \\ 0.828, & n = 0.9; \\ 0.855, & n = 1. \end{cases} \tag{67}$$

Again, we have assumed $s = 1$ for all these cases.

## 5.2. The Median Observed Value of the Vacuum Energy Density in a COBE-Normalized CDM Universe

We can calculate the median value of the vacuum energy density observed in all subuniverses, $(\rho_V)_{1/2}$, by using the solutions for $\alpha_{1/2} \equiv (\rho_V)_{1/2} / <\rho_V>$ given in equation (32), multiplied by $<\rho_V>$ as calculated already from equation (60). The latter depends upon the local values adopted for $\lambda_0^*$ and $h^*$, used in calculating the variance $\sigma^2$. Since the value of $\alpha_{1/2}$ is, for a given $s$, just a number, independent of $\sigma$, it is sufficient for us to write $(\rho_V)_{1/2} = \alpha_{1/2}\langle\rho_V\rangle$ (as for instance $\alpha_{1/2} = 0.43411$ for $s = 1$) and use our previous results for $<\rho_V>$. For convenience, we have plotted $\rho_V^*/(\rho_V)_{1/2}$ versus $\lambda_0^*$ and versus $\lambda_0^*/\Omega_0^*$, for $h^* = 0.5$ and $1$, $R_G = 0.01, 1, 2,$ and $3\,\mathrm{Mpc}$, and $n = 1, 0.9, 0.8$ in Figure 9.

*For what values of $\lambda_0^*$ and $h^*$ is the vacuum energy density of our own subuniverse equal to the median value $(\rho_V)_{1/2}$ observed in all subuniverses?* For the particular values of $R_G$ and $h^*$ assumed in plotting the curves in Figure 9, the intersection of the horizontal dashed lines with the curves indicates the values for which $\rho_V^* = (\rho_V)_{1/2}$. To answer this question more generally as a function of $R_G$ and $h^*$, we must solve the implicit equation $(\rho_V)_{1/2} = \rho_V^*$. This is similar to our previous implicit equation $<\rho_V> = \rho_V^*$, except replaced by the equation $\alpha_{1/2} <\rho_V> = \rho_V^*$, where $\alpha_{1/2}$ is a constant for a given $s$ (and only weakly depends on $s$). The results are shown in Table 3 for $n = 1$ and $s = 1$. We find that for $R_G = 1\,\mathrm{Mpc}$, our own subuniverse has a vacuum energy density equal to that of the median if

$$\frac{\rho_V^*}{\rho_0^*} \cong \begin{cases} 3.4, \\ 6.1, \\ 10.3, \end{cases} \qquad \lambda_0^* \cong \begin{cases} 0.77, & h^* = 0.5\ ; \\ 0.86, & h^* = 0.7\ ; \\ 0.91, & h^* = 1\ ; \end{cases} \tag{68}$$

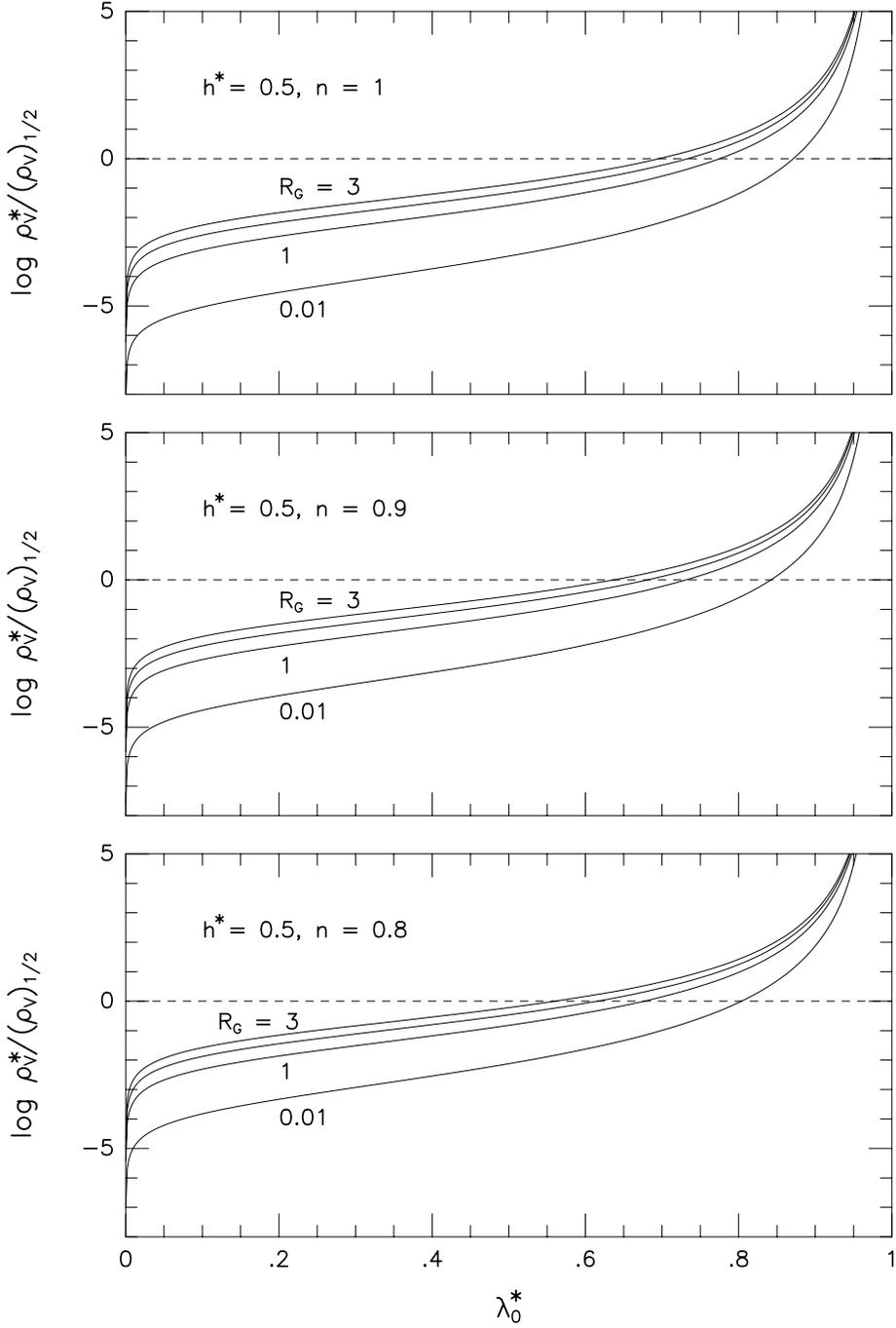

Fig. 9.— (a) The vacuum energy density adopted for our own subuniverse, $\rho_V^*$, is plotted in units of the median value of the vacuum energy density observed in all subuniverses, $(\rho_V)_{1/2}$, versus $\lambda_0^*$, the cosmological constant in our own subuniverse, for local Hubble constant $h^* = 0.5$, for a COBE-normalized, flat CDM model with primordial spectral index $n = 1$ (top panel), 0.9 (middle panel), and 0.8 (bottom panel), assuming shape parameter $s = 1$. Intersection of the dashed horizontal lines at $\rho_V^*/(\rho_V)_{1/2} = 1$ with the various curves give the solution of the implicit equation obtained by setting $(\rho_V)_{1/2} = \rho_V^*$ in equations (27), (28), and (59); (b) Same as Fig. 9(a), except plotted versus $\lambda_0^*/\Omega_0^*$; (c) Same as Fig. 9(a), except for $h^* = 1$; (d) Same as Fig. 9(b), except for $h^* = 1$.

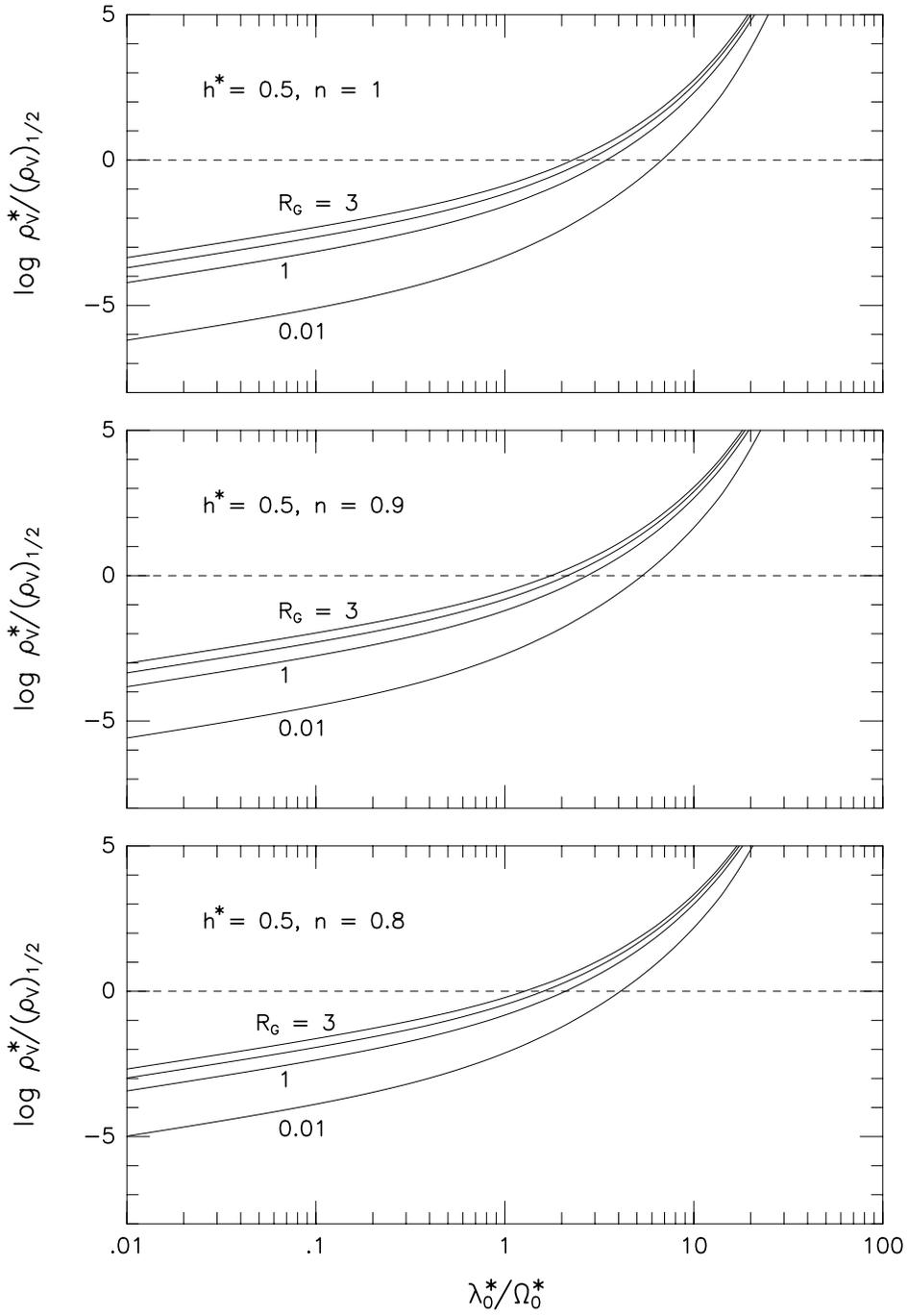

Fig. 9b

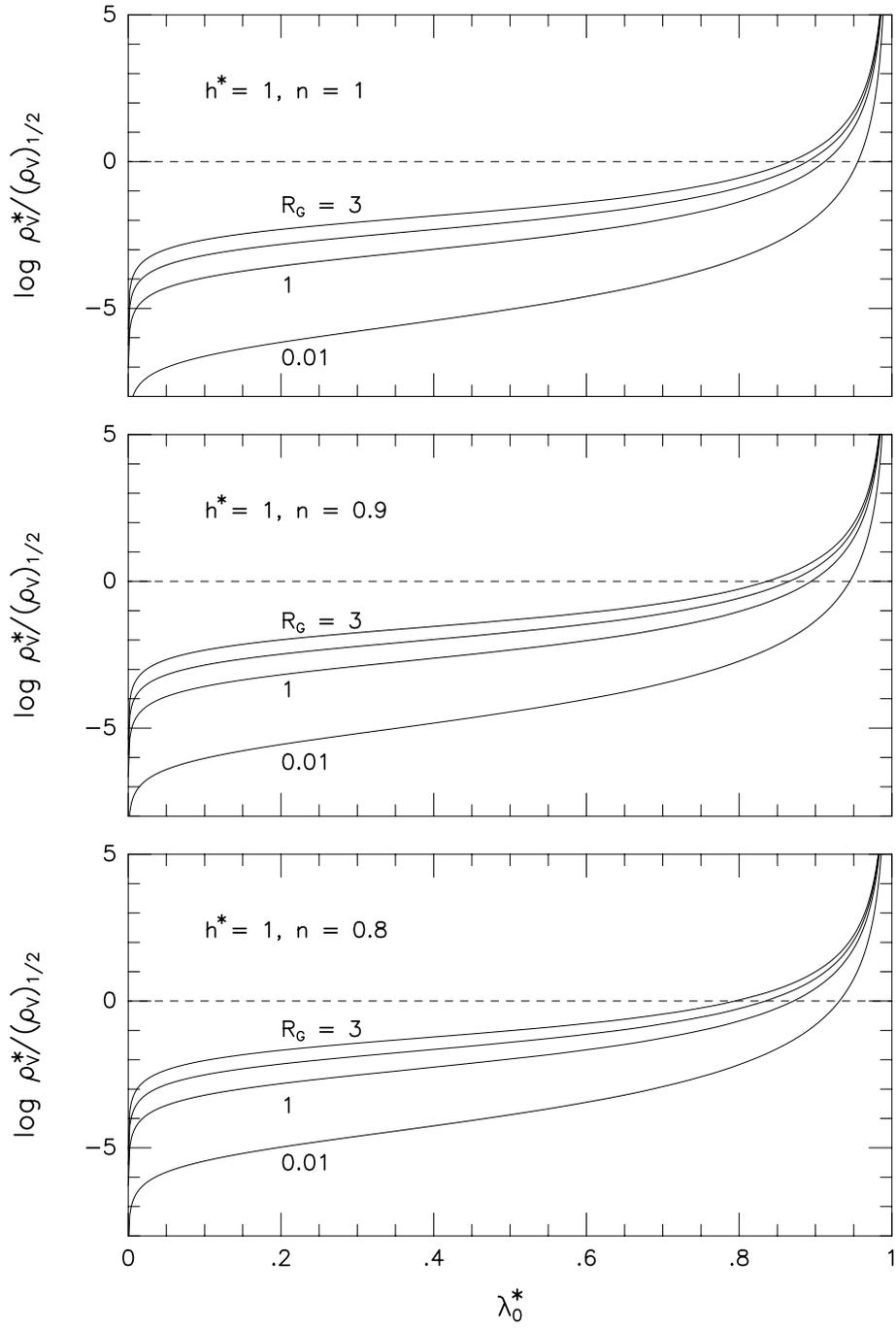

Fig. 9c

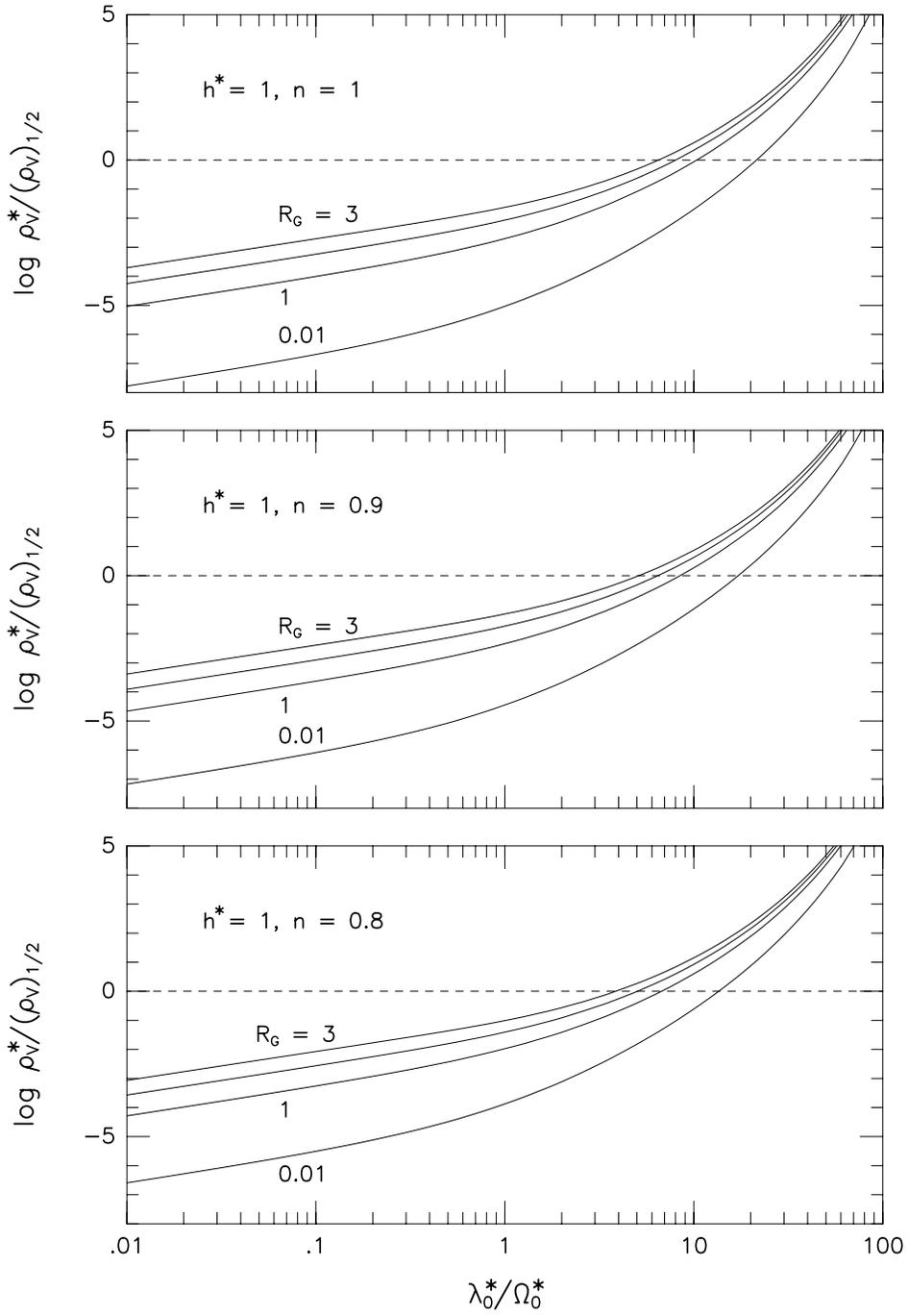

Fig. 9d

Table 3.   LOCAL VACUUM ENERGY DENSITIES WHICH EQUAL THE GLOBAL MEDIAN $(\rho_V^* = (\rho_V)_{1/2})$[1]

| $R_{\rm G}$ (Mpc)$^2$ | $h^* = 0.5$ | | | $h^* = 0.7$ | | | $h^* = 1$ | | |
|---|---|---|---|---|---|---|---|---|---|
| | $\lambda_0^*$ | $\Omega_0^*$ | $\lambda_0^*/\Omega_0^*$ | $\lambda_0^*$ | $\Omega_0^*$ | $\lambda_0^*/\Omega_0^*$ | $\lambda_0^*$ | $\Omega_0^*$ | $\lambda_0^*/\Omega_0^*$ |
| 0.003 | 0.8795 | 0.1205 | 7.2988 | 0.9301 | 0.0699 | 13.306 | 0.9592 | 0.0408 | 23.510 |
| 0.010 | 0.8709 | 0.1291 | 6.7459 | 0.9239 | 0.0761 | 12.141 | 0.9553 | 0.0447 | 21.371 |
| 0.030 | 0.8592 | 0.1408 | 6.1023 | 0.9164 | 0.0836 | 10.962 | 0.9504 | 0.0496 | 19.161 |
| 0.100 | 0.8414 | 0.1586 | 5.3052 | 0.9049 | 0.0951 | 9.5152 | 0.9429 | 0.0571 | 16.513 |
| 0.300 | 0.8175 | 0.1825 | 4.4795 | 0.8891 | 0.1109 | 8.0171 | 0.9324 | 0.0676 | 13.793 |
| 1.000 | 0.7740 | 0.2260 | 3.4248 | 0.8590 | 0.1410 | 6.0922 | 0.9115 | 0.0885 | 10.299 |
| 2.000 | 0.7319 | 0.2681 | 2.7300 | 0.8273 | 0.1727 | 4.7904 | 0.8884 | 0.1116 | 7.9606 |
| 3.000 | 0.6940 | 0.3060 | 2.2680 | 0.7988 | 0.2012 | 3.9702 | 0.8659 | 0.1341 | 6.4571 |
| 6.000 | 0.5877 | 0.4123 | 1.4254 | 0.7036 | 0.2964 | 2.3738 | 0.7761 | 0.2239 | 3.4663 |
| 10.000 | 0.4279 | 0.5721 | 0.7480 | 0.5080 | 0.4920 | 1.0325 | 0.4584 | 0.5416 | 0.8464 |
| 15.000 | 0.2231 | 0.7769 | 0.2872 | 0.1889 | 0.8111 | 0.2389 | 0.0760 | 0.9240 | 0.0823 |
| 20.000 | 0.0972 | 0.9028 | 0.1077 | 0.0582 | 0.9418 | 0.0618 | 0.0179 | 0.9821 | 0.0182 |

[1] For Harrison-Zel'dovich scale-invariant primordial spectrum ($n = 1$), and $s = V/U = 1$.

[2] Present units

while if $R_G = 2\,\mathrm{Mpc}$, these values shift to

$$\frac{\rho_V^*}{\rho_0^*} \cong \begin{cases} 2.7\,, \\ 4.8\,, \\ 8.0\,, \end{cases} \qquad \lambda_0^* \cong \begin{cases} 0.73\,, & h^* = 0.5\;; \\ 0.83\,, & h^* = 0.7\;; \\ 0.89\,, & h^* = 1\,. \end{cases} \qquad (69)$$

If $R_G = 3\,\mathrm{Mpc}$, then our subuniverse has the median observed value of $\rho_V$ if

$$\frac{\rho_V^*}{\rho_0^*} \cong \begin{cases} 2.3\,, \\ 4.0\,, \\ 6.5\,, \end{cases} \qquad \lambda_0^* \cong \begin{cases} 0.69\,, & h^* = 0.5\;; \\ 0.80\,, & h^* = 0.7\;; \\ 0.87\,, & h^* = 1\,. \end{cases} \qquad (70)$$

If, on the other hand, $R_G = 0.01\,\mathrm{Mpc}$, then this happens if

$$\frac{\rho_V^*}{\rho_0^*} \cong \begin{cases} 6.7\,, \\ 12.1\,, \\ 21.3\,, \end{cases} \qquad \lambda_0^* \cong \begin{cases} 0.87\,, & h^* = 0.5\;; \\ 0.92\,, & h^* = 0.7\;; \\ 0.96\,, & h^* = 1\,. \end{cases} \qquad (71)$$

Here again, the effect of a tilt to $n < 1$ is to decrease the values of $\rho_V^*$ and of $\lambda_0^*$ corresponding to the median $\rho_V$ somewhat. For example, for $R_G = 1\,\mathrm{Mpc}$ and $h^* = 0.7$, our own subuniverse has the median value of $\rho_V$ (assuming $s = 1$) for

$$\lambda_0^* \cong \begin{cases} 0.80\,, & n = 0.8\;; \\ 0.83\,, & n = 0.9\;; \\ 0.86\,, & n = 1\,. \end{cases} \qquad (72)$$

For $R_G = 2\,\mathrm{Mpc}$ and $h^* = 0.7$, this becomes

$$\lambda_0^* \cong \begin{cases} 0.75\,, & n = 0.8\;; \\ 0.79\,, & n = 0.9\;; \\ 0.83\,, & n = 1\,. \end{cases} \qquad (73)$$

### 5.3. Observational Constraints on the Cosmological Constant in Our Own Subuniverse

Ostriker & Steinhardt (1995) have argued that the apparent discrepancies that had previously been identified between observations and the predictions of the standard CDM model (i.e. cold dark matter in a flat universe with zero cosmological constant and the scale-invariant Harrison-Zel'dovich primordial power spectrum) can be reconciled if the standard CDM model is modified to admit a nonzero cosmological constant roughly in the range $\lambda_0^* = 0.65 \pm 0.1$. We have reproduced the observational and theoretical constraints that led

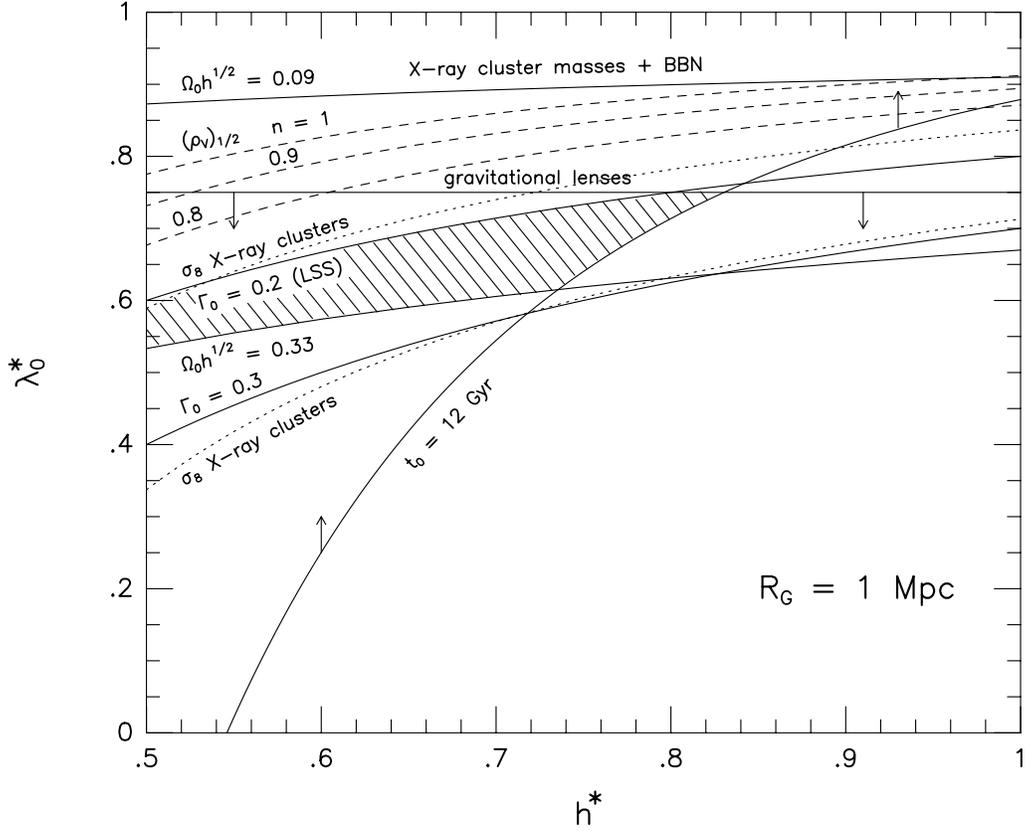

Fig. 10.— (a) Observational Constraints on $\lambda_0^*$ and $h^*$ for own subuniverse. Curves labelled "LSS" and "$\Gamma_0 = 0.2$" or "$\Gamma_0 = 0.3$" bound the region allowed by the constraint $\Gamma_0^* = \Omega_0^* h^* = 0.25 \pm 0.05$ derived by matching the spatial and angular correlation statistics from galaxy surveys with the theoretical predictions of the large-scale clustering of galaxies in a COBE-normalized, flat CDM model with primordial power spectrum index $n = 1$. The curves labelled "$\sigma_8$ X-ray clusters" bound the values of $\lambda_0^*$ and $h^*$ which make this CDM model satisfy the constraint on the present space density of X-ray clusters. The curve labelled "$t_0 = 12\,\mathrm{Gyr}$" indicates the lower limit which makes the age of the universe at least as large as current estimates of the minimum age of globular clusters. The curves labelled "$\Omega_0 h^{1/2}$" indicate the boundaries defined by the X-ray-measured total and baryonic masses of clusters of galaxies, together with the big bang nucleosynthesis limits on the baryon mean density and the assumption that the ratio of baryon to total mass inside each cluster equals the ratio of universal mean values. The curve labelled "gravitational lenses" indicates the upper limit imposed by counts of quasars lensed by intervening galaxies. The dashed curves labelled "$(\rho_V)_{1/2}$" are the values for which our own subuniverse has the median value of $\rho_V$ for all subuniverses, if $R_G = 1\,\mathrm{Mpc}$, if $n = 1$ (top dashed curves), 0.9 (middle dashed curve), or 0.8 (bottom dashed curve). (b) Same as Fig. 10(a), except $R_G = 2\,\mathrm{Mpc}$. (c) Same as Fig. 10(a), except $R_G = 3\,\mathrm{Mpc}$.

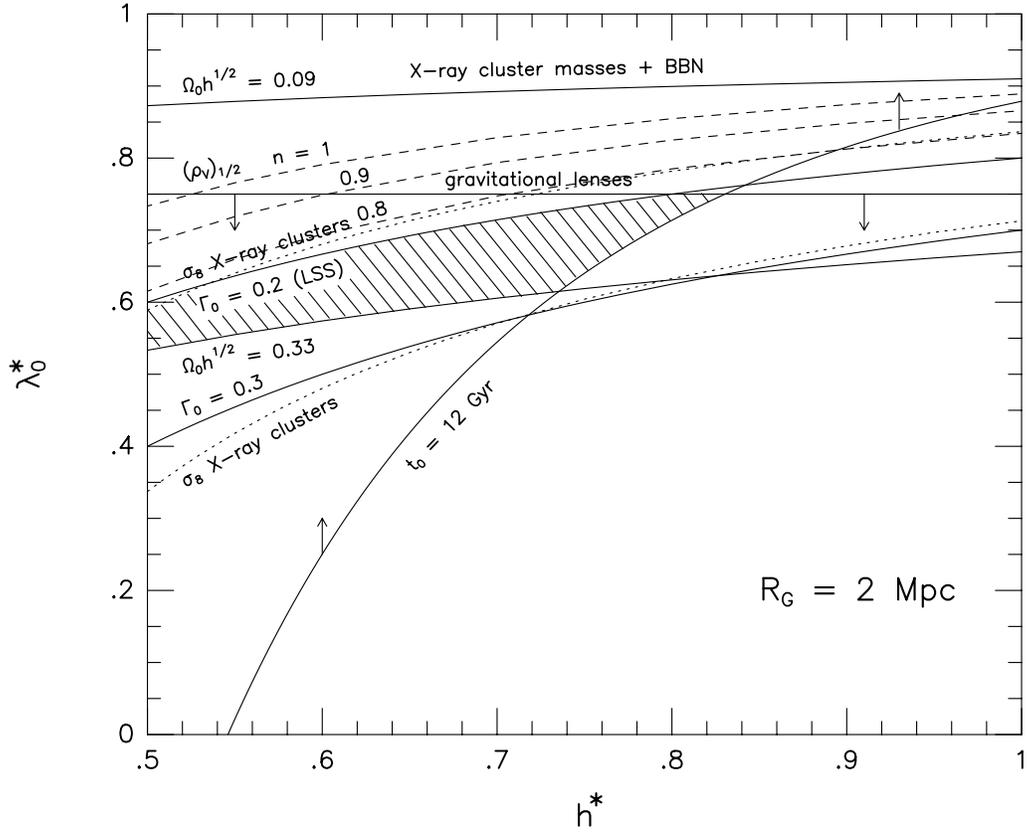

Fig. 10b

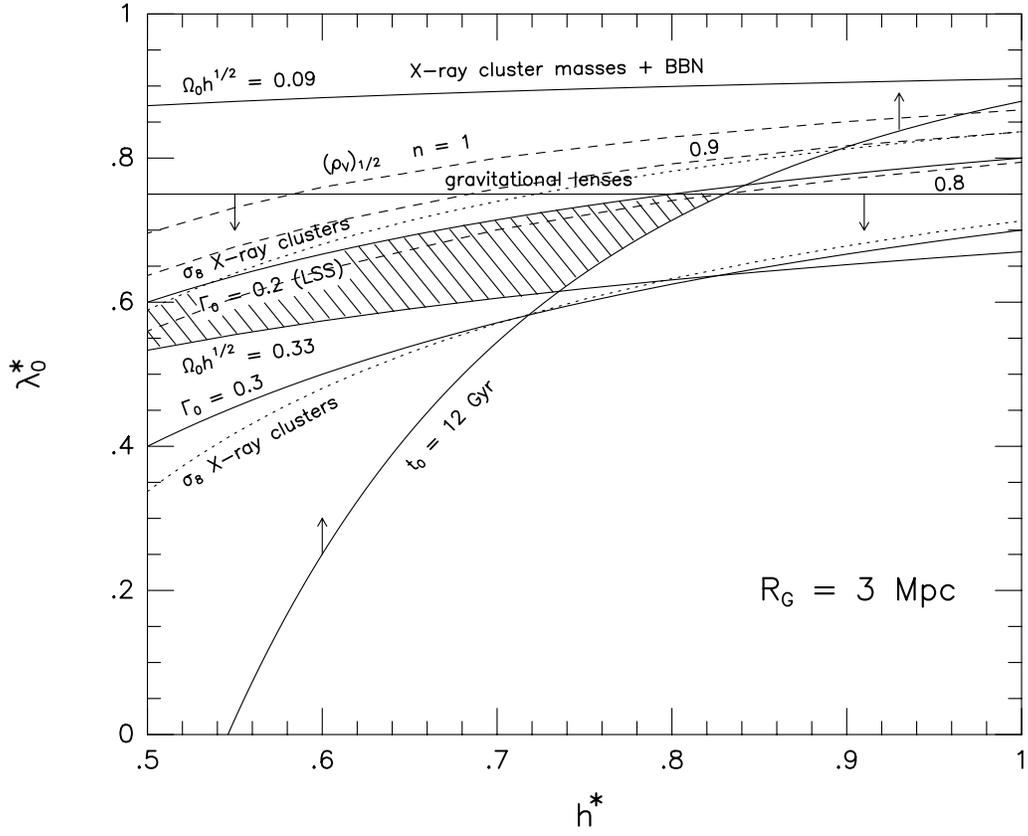

Fig. 10c

Ostriker & Steinhardt (1995) to this conclusion, plotted here in Figure 10 as a series of upper and lower bounds in the $(\lambda_0^*, h^*)$-plane. The constraint curves plotted in Figure 10 are based on Ostriker & Steinhardt (1995) and references therein, with the following additions.

The data on the large-scale clustering of galaxies from galaxy surveys constrain the flat, CDM model by requiring that the spatial and angular correlation statistics of the observed galaxies in our local universe at the present epoch agree with the predictions of structure formation by gravitational instability in the CDM model. This leads to upper and lower bounds on the so-called "shape parameter" $\Gamma_0 = \Omega_0 h$, given by $\Gamma_0^* = 0.25 \pm 0.05$ (assuming $n = 1$) (see curves labelled "$\Gamma_0$" and "LSS" in Fig. 10). A similar constraint results from the requirement that the CDM model reproduce the observed space density and luminosity function of X-ray clusters in the present universe. We plot this constraint separately from that of the shape parameter $\Gamma_0$ which Ostriker & Steinhardt (1995) plotted. This X-ray cluster abundance constraint is expressed by Viana & Liddle (1996) as bounds on the rms density fluctuation $\sigma_8^*(z = 0)$ for a smoothing radius (in present units) given by $R_{\mathrm{TH}} \approx 8h^{-1}\mathrm{Mpc}$. Assuming $n = 1$, these bounds are given by

$$\sigma_8^* = \left[ 0.6\Omega_0^{*-(0.59 - 0.016\Omega_0^* + 0.06\Omega_0^{*2})} \right]^{+32\phi(\Omega_0^*)\%}_{-24\phi(\Omega_0^*)\%}, \tag{74}$$

where

$$\phi(\Omega_0^*) \equiv (\Omega_0^*)^{0.26\log_{10}\Omega_0^*}. \tag{75}$$

This amounts to a constraint on $\lambda_0^*$ and $h^*$ which is similar to that from the correlation statistics. (The above-mentioned bounds from the statistics of large-scale structure ("LSS") in the galaxy distribution also refer to the part of the CDM power spectrum at wavelengths $\lambda \gtrsim 8h^{-1}\mathrm{Mpc}$.)

Estimates of the total masses and baryonic mass fractions of individual clusters of galaxies, derived by fitting the X-ray surface brightness profiles of each cluster and assuming the cluster intergalactic medium is an isothermal sphere in hydrostatic equilibrium with a virialized cluster gravitational potential, yield another pair of bounds. If the assumption is further made that the ratio of baryonic mass to total mass of each cluster is equal to the universal mean ratio, $\Omega_{B0}/\Omega_0$, then a comparison of this X-ray-estimated ratio with the constraints on $\Omega_{B0}$ from standard big bang nucleosynthesis and the observed light element abundances (i.e. $0.01 \lesssim \Omega_{B0}^*(h^*)^2 \lesssim 0.02$; Copi, Schramm, & Turner 1995) implies a constraint on the total density parameter given by $0.09 \lesssim \Omega_0^*(h^*)^{1/2} \lesssim 0.33$. The curves labelled "$\Omega_0 h^{1/2}$" and "X-ray cluster masses + big bang nucleosynthesis" in Figure 10 indicate the bounds on $\lambda_0^*$ and $h^*$ which result from this argument. Some recent numerical gas dynamical simulations of cluster formation in the flat, matter-dominated CDM model suggest that the upper bound on $\Omega_0 h^{1/2}$ which results from the high values estimated for cluster baryonic mass fraction by the equilibrium model described above may be too low (e.g. Bartelmann & Steinmetz 1996; Martel, Shapiro, & Valinia 1996; Valinia 1996). The simulated clusters, when properly resolved, are often found to be comprised of subclusters in the act of merging and, together with projection effects, this can cause an observer who uses the assumption of isothermal spheres in hydrostatic equilibrium to underestimate the total mass and overestimate the baryon fraction.

The estimated minimum age of globular clusters derived by comparison of theoretical models of stellar evolution with observed globular cluster H-R diagrams is about $12 \times 10^9$ years. This leads to a lower limit to $\lambda_0^*$ for each $h^*$ based on the requirement that the age of our universe exceeds this estimate of the minimum age of globular cluster stars (see curve in Fig. 10 labelled "$t_0 = 12\,\mathrm{Gyr}$").

An upper bound to $\lambda_0^*$ results from the comparison of the statistics of quasars observed to be gravitationally lensed by intervening galaxies with the predictions of flat cosmological models with a nonzero cosmological constant. A flat cosmology with cosmological constant tends to produce more gravitationally lensed quasars than does such a cosmology with zero cosmological constant. The resulting limit was quoted by Ostriker & Steinhardt as $\lambda_0^* < 0.75$. More recently, Kochanek (1996) has argued for a somewhat tighter limit. However, limited observational data and the possibility that evolution effects on the population of lensing galaxies have not been properly taken into account suggest that this limit is still uncertain. We plot the upper bound quoted by Ostriker & Steinhardt (1995) in Figure 10, but caution that this limit is still in flux and that the future change can be *up or down*.

Finally, an independent constraint was recently derived by Perlmutter et al. (1996), using observations of Type Ia supernovae to infer the relationship between redshift and distance. Their result is consistent with the limit $\lambda_0^* < 0.5$. However, this interesting approach is too preliminary to be considered reliable at this time. We have not included it in Figure 10.

## 5.4. How Likely is the Value of the Cosmological Constant Observed in Our Subuniverse?

Along with the observational constraints discussed in the previous subsection, we have plotted in Figure 10 the values of the cosmological constant $\lambda_0^*$ such that our own universe has the median value observed for all subuniverses for different values of $h^*$ and for $n = 1$, 0.9, and 0.8, for $R_G = 1$, 2, and $3\,\mathrm{Mpc}$ and $s = 1$. It is apparent from Figure 10 that if the universe is flat and the inflationary CDM model applies, then the range of values of the cosmological constant allowed for our own subuniverse is somewhat below the median, but not far from it.

This statement is sensitive to the choice of the smoothing scale $R_G$. As discussed in Section 4.1, this scale is intended to reflect the minimum size of fluctuations which are responsible for forming observers in any subuniverse. The value $R_G = 1 \, \text{Mpc}$ was chosen to correspond to fluctuations which encompass roughly the mass of the bright inner part of an $L_*$-galaxy. If, instead, observers are formed if even a much smaller, globular-cluster-size object collapses out of the background, then the global median value of $\rho_V$ would be larger since the variance of the density fluctuations at recombination filtered on this smaller scale is larger. In that case, the median $(\rho_V)_{1/2}$ is on the high side, somewhat further from the range in the $(\lambda_0^*, h^*)$-plane allowed by observations of our own universe. Conversely, if observers are formed only if fluctuations $larger$ than 1 Mpc collapse out, then the global median $(\rho_V)_{1/2}$ will be smaller [since $\sigma(z_{rec})$ is smaller if $R_G > 1 \, \text{Mpc}$ than if $R_G = 1 \, \text{Mpc}$] and the match between the allowed range in the $(\lambda_0^*, h^*)$-plane and $(\rho_V)_{1/2}$ improves. We illustrate this by plotting the median value curves for $R_G = 2$ and 3 Mpc along with the observational constraints, in Figures 10(b) and (c).

These results also depend weakly on the value adopted for $\Omega_{B0}^*$. A small increase of $\Omega_{B0}^*$ has the effect of decreasing $\Gamma^*$ and, with it, the value of $\sigma(z_{rec})$. This, too, would drive the global median $(\rho_V)_{1/2}$ downward for given values of $R_G$ and $h^*$. We chose $\Omega_{B0}^* = 0.015(h^*)^{-2}$ here, but the big bang nucleosynthesis limits allow a value $\Omega_{B0}^* = 0.02(h^*)^{-2}$ and uncertainties exist even in that upper limit.

Even though $\rho_V$ in our subuniverse seems to be below the median, the allowed range in the $(\lambda_0^*, h^*)$- plane in Figure 10 includes values that are reasonably likely. To see this, it is useful to recast the curves of Figure 10 as curves in the $(\alpha^*, h^*)$-plane, where $\alpha \equiv \rho_V / <\rho_V>$. For each point $(\lambda_0^*, h^*)$ on a curve in Figure 10 we can use the results of Sections 3 and 4 to compute $<\rho_V>$, and hence the ratio $\alpha^* \equiv \rho_V^* / <\rho_V>$. We plot these constraint curves in the $(\alpha^*, h^*)$-plane in Figure 11. Inspection of this figure shows that the observationally allowed values of $\alpha^*$ are between 0.01 and 0.04 for $R_G = 1$ Mpc, between 0.015 and 0.1 for $R_G = 2$ Mpc, and between 0.03 and 0.15 for $R_G = 3$ Mpc. For all of these values of $R_G$, there is an appreciable overlap between these observationally allowed values, and the range from 0.02 to 3.7 which was found (for $s = 1$) in equations (33) and (34) to be anthropically likely.

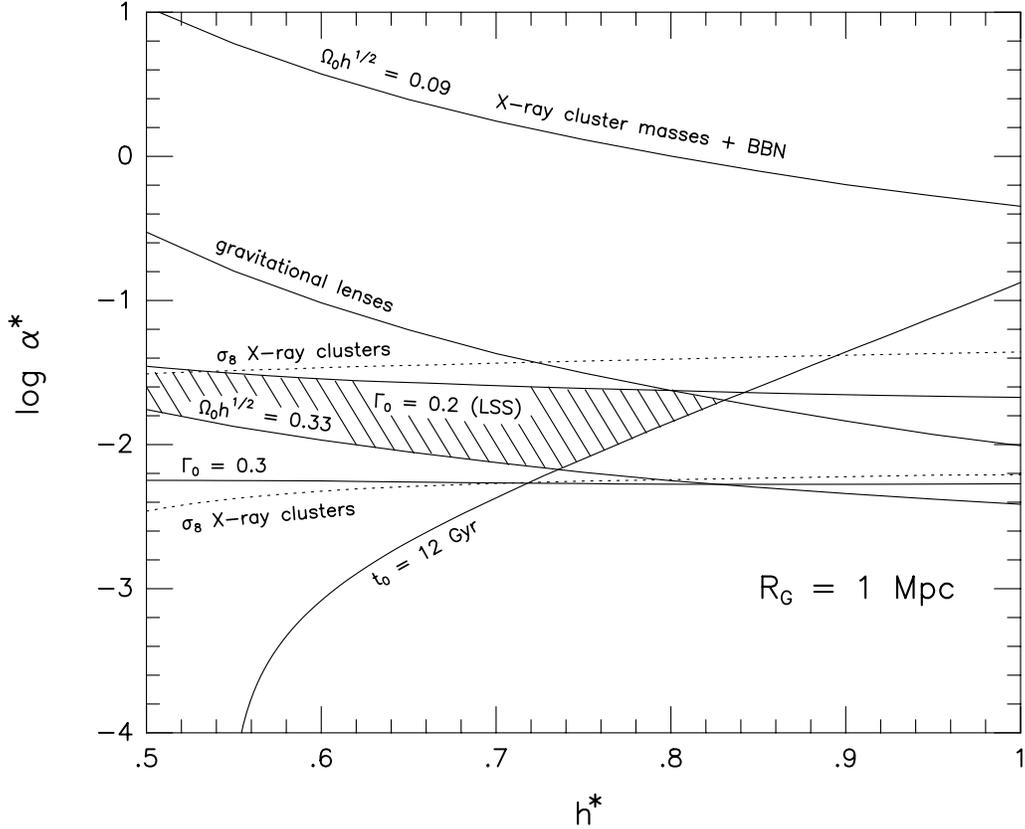

Fig. 11.— (a) Observational Constraints of Fig. 10 are plotted instead as curves in the $(\alpha^*, h^*)$-plane, where $\alpha^* \equiv \rho_V^* / <\rho_V>$, and $<\rho_V>$ is evaluated by assuming COBE-normalized density fluctuations for the flat CDM model with $n = 1$, as inferred by adopting the value $\rho_V = \rho_V^*$, assuming $R_G = 1\,\mathrm{Mpc}$ and $s = 1$. Curves are labelled just as in Figure 10. (b) Same as Fig. 11(a), except $R_G = 2\,\mathrm{Mpc}$. (c) Same as Fig. 11(a), except $R_G = 3\,\mathrm{Mpc}$.

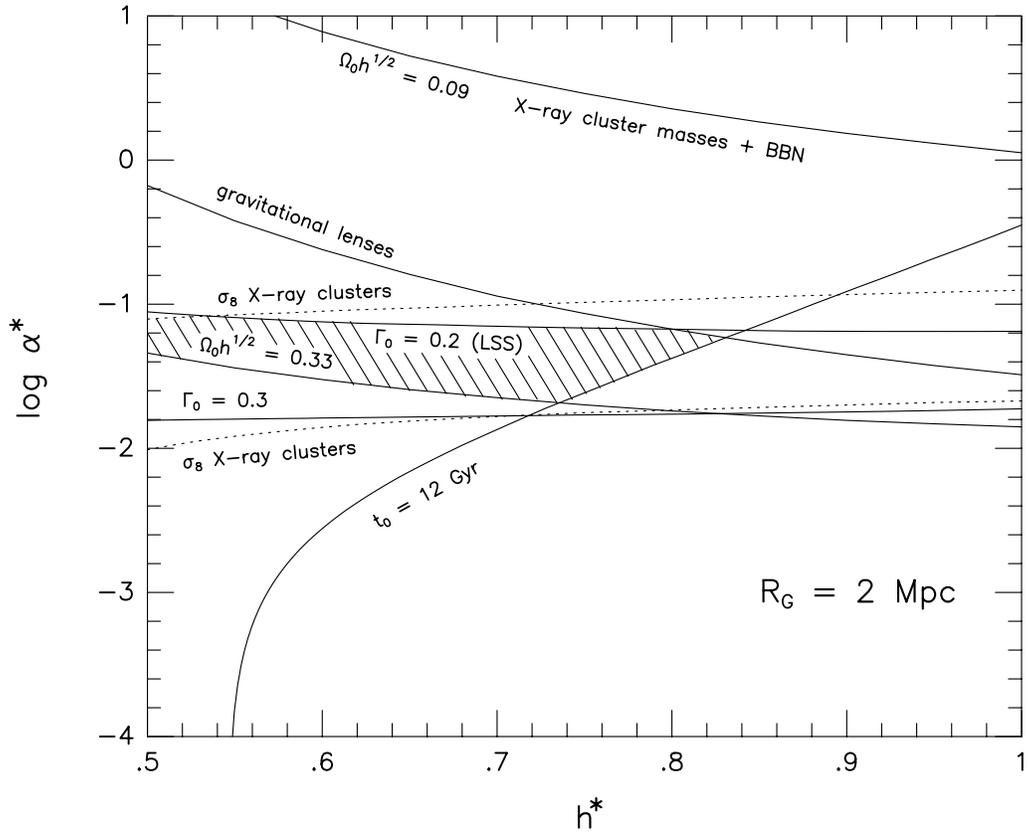

Fig. 11b

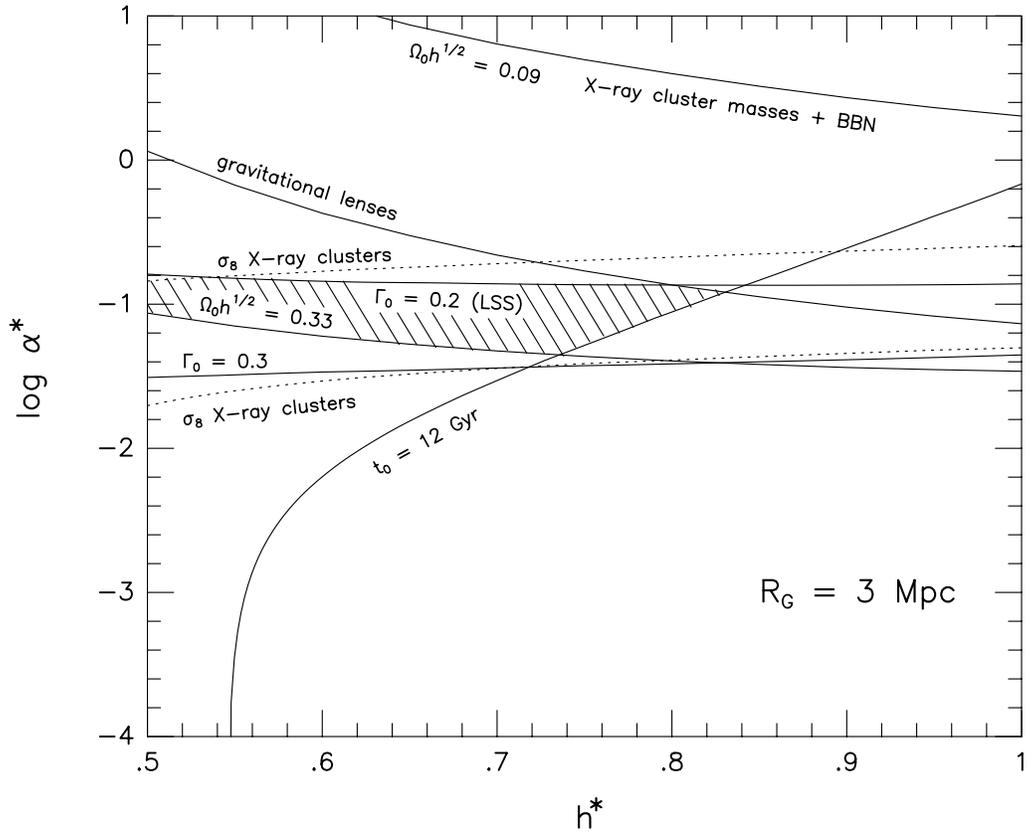

Fig. 11c

## 6.  SUMMARY AND CONCLUSION

The range of values of the cosmological constant that allow life to arise is so narrow, that within that range, we can assume that the *a priori* probability distribution of values of the cosmological constant is constant. The probability that a particular value of the cosmological constant is observed in our universe is then proportional to the number of observers who might measure that value. That abundance is, in turn, proportional to the fraction of matter which eventually collapses out of the background into gravitationally bound concentrations large enough to initiate star formation and retain heavy elements, presumed to be prerequisites for the origin of planets and intelligent life. We have derived an analytical estimate for this "collapsed fraction" based upon a simple, pressure-free, spherically symmetric, nonlinear model for the growth of density fluctuations in a flat universe with arbitrary value of the vacuum energy density $\rho_V$, applied in a statistical way to a distribution of cosmological density fluctuations. We have evaluated the resulting probability distribution for the observed values of $\rho_V$ for density fluctuations which are Gaussian random and of linear amplitude at recombination. We find that the probability distribution in that case is a unique function of $\rho_V / < \rho_V >$, where $< \rho_V >$ is the average of the observed values of $\rho_V$ over all subuniverses, and a shape parameter $s$ which characterizes the variation of density with radius within a given density fluctuation. The dependence on the parameter $s$ is, moreover, found to be very weak. The values of $\rho_V$ (such as the median value) at which the integrated probability distribution takes any definite values are simply proportional to the mean value $< \rho_V >$ for a given $s$, with proportionality constants that are fairly insensitive to the value of $s$, and for a given value of $s$, these values of $\rho_V$ are, in turn, exactly proportional to the quantity $\sigma^3 \bar{\rho}$ evaluated at recombination, where $\sigma^2$ is the variance of the density fluctuations and $\bar{\rho}$ is the cosmic mean matter density. The dependence of this proportionality constant on the value of $s$ is, once again, quite weak. The quantity $\sigma^3 \bar{\rho}$ is assumed to be common to all subuniverses in which life can arise, because in these subuniverses $\rho_V$ is negligible at and before recombination.

Presumably it will some day be possible to calculate $\sigma^3 \bar{\rho}$ unambiguously by using astronomical observations to measure the density fluctuations in our own universe. While this is not yet possible with great certainty, current measurements of anisotropy at large angles in the cosmic microwave background by the COBE DMR experiment represent substantial progress toward this goal. In particular, the detected cosmic microwave background anisotropy fixes the fluctuation amplitude at long wavelengths and is consistent with Gaussian random density fluctuations in a flat universe, with the scale- invariant primordial power spectrum, $P(k) \propto k^n$ with $n \approx 1$, as expected from inflationary cosmology. Unfortunately, the precise interpretation of these anisotropy measurements is, itself, dependent upon our uncertain knowledge of the values of $\lambda_0$ and $h$ for our observed universe. In addition, the variance $\sigma^2$ which is relevant to our anthropic probability calculation is that for the density fluctuations after the density has been smoothed over some length scale $R_G$ so as to eliminate fluctuations which are too small to contribute to the formation of astronomers, and $R_G$ is orders of magnitudes smaller than the long wavelengths at which the fluctuations are constrained directly by the COBE DMR anisotropy measurements. Unfortunately, therefore, neither the value of $R_G$ nor the amplitude of the density fluctuations at wavelengths close to $R_G$ are precisely determined at this time. In the meantime, in order to specify $\sigma^3\bar{\rho}$, we have here adopted the CDM model for the growth of a scale-invariant primordial spectrum of Gaussian random density fluctuations in a flat universe with nonzero cosmological constant, with an amplitude set by the COBE DMR data, and parameterized our results in terms of the unknown filter scale $R_G$ and power spectrum index $n$.

Our results are encouraging from the point of view of the anthropic hypothesis. Although the range of observationally favored values of $\rho_V$ is somewhat less than the median value of $\rho_V$ for all subuniverses, it has a significant overlap with the range of values between the 5th and 95th percentile for all subuniverses.

In short, these results explain why there might be a nonzero, but small value of $\rho_V$ in our universe, if all values of $\rho_V$ are otherwise equally likely to occur. They show that a range of values close to those favored by current observations are reasonably probable, while values which are orders of magnitude smaller or larger are extremely improbable for us to observe. This is the essential ingredient in the anthropic explanation for the observed value of $\lambda_0$ in our own universe.

With a continued improvement in measurements of $\rho_V$ and the spectrum of fluctuations at recombination, it may turn out that the actual value of $\rho_V$ in our subuniverse is more than two orders of magnitude less than the average. In this case we would have to conclude that the anthropic arguments used here do not explain the smallness of the cosmological constant. Unfortunately, the converse is not possible; observation of a value of $\rho_V$ that is anthropically likely would support the idea that there is a diversity of possible $\rho_V$ values, but since we only observe one subuniverse, astronomical observation alone cannot confirm this idea. Ultimately this issue will have to be settled by advances in fundamental physics, which we hope will tell us whether in fact it is correct that there are many subuniverses with different values of the cosmological constant. If this is not correct, then there is no justification for the anthropic reasoning used here, while if it is correct, then these anthropic arguments are just common sense.

This research was supported in part by NASA Grant NAG5-2785, NSF Grants ASC